%% file: sample-sigconf-authordraft.tex
\PassOptionsToPackage{table,xcdraw}{xcolor}
\documentclass[sigconf]{acmart}

\renewcommand\footnotetextcopyrightpermission[1]{}  % keep this too
\setcopyright{none}                % removes copyright/permission block entirely
\acmDOI{}                          % blank out the DOI placeholder
\acmISBN{}                         % blank out the ISBN placeholder
\acmConference[arXiv preprint]{2503.16640}{2025}{Online}
\renewcommand\footnotetextcopyrightpermission[1]{}
\AtBeginDocument{%
  }

\usepackage[table,xcdraw]{xcolor}
\usepackage[framemethod=TikZ]{mdframed}
\usepackage{float}
\usepackage{subcaption}
\usepackage{listings}
\usepackage{stackengine} 
\usepackage{subcaption}
\usepackage{adjustbox}
\definecolor{cyan1}{RGB}{0,255,255}
\definecolor{codegreen}{rgb}{0,0.6,0}
\definecolor{codegray}{rgb}{0.5,0.5,0.5}
\definecolor{codepurple}{rgb}{.58,0,0.82}
\definecolor{backcolour}{rgb}{0.95,0.95,0.92}
\usepackage{comment}
\usepackage{array,multirow}
\usepackage{rotating}
\usepackage{url}
\usepackage{todonotes}
\usepackage{hyperref}
\usepackage[nameinlink]{cleveref}
\usepackage{cleveref}
\definecolor{lightblue}{RGB}{200, 220, 255}
\definecolor{vsdarkbackground}{RGB}{30, 30, 30}
\definecolor{vsgreybackground}{RGB}{77,74,60}
\definecolor{vsdarkblue}{RGB}{94, 155, 255}
\definecolor{vsdarkgreen}{RGB}{120, 195, 80}
\definecolor{vsdarkred}{RGB}{255, 85, 85}
\definecolor{vsdarkpurple}{RGB}{177, 97, 211}
\definecolor{vsdarkgray}{RGB}{155, 155, 155}
\definecolor{lightblue}{RGB}{200, 220, 255}
\definecolor{lightgrey}{RGB}{230, 230, 230}
\definecolor{lightorange}{RGB}{255, 201, 102}
\definecolor{lightyellow}{RGB}{255, 255, 0}
\definecolor{loggray}{RGB}{160,160,160}
\definecolor{codegray}{rgb}{0.5,0.5,0.5}
\definecolor{codepurple}{rgb}{0.58,0,0.82}
\definecolor{backcolour}{rgb}{0.95,0.95,0.92}
\definecolor{lightred}{RGB}{255, 110, 110}
\definecolor{lightgreen}{RGB}{181, 230, 29}
\definecolor{codegreen}{rgb}{0,0.6,0}
\newcommand{\eb}[1]{\todo[inline,size=small,color=yellow]{EB: #1}}
\newcommand{\mk}[1]{\todo[inline,size=small,color=lightblue]{MK: #1}}
\newcommand{\ms}[1]{\todo[inline,size=small,color=orange]{MS: #1}}

\newcommand{\stilltodo}[1]{\textcolor{red}{#1}}

\makeatother

\lstdefinestyle{mystyle}{   
    commentstyle=\color{codegreen},
    keywordstyle=\color{magenta},
    numberstyle=\tiny\color{codegray},
    stringstyle=\color{codepurple},
    basicstyle=\ttfamily\footnotesize,
    breakatwhitespace=false,         
    breaklines=true,                 
    captionpos=b,   
    frame=single,
    keepspaces=true,                 
    numbers=left,                    
    numbersep=5pt,                  
    showspaces=false,                
    showstringspaces=false,
    showtabs=false,                  
    tabsize=2
}

\newcommand{\tikzcircleblack}[2][black,fill=black]{\tikz[baseline=-0.5ex]\draw[#1,radius=#2] (0,0) circle ;}%

\newcommand{\tikzcirclewhite}[2][black,fill=white]{\tikz[baseline=-0.5ex]\draw[#1,radius=#2] (0,0) circle ;}%
\begin{document}

%%
%% The "title" command has an optional parameter,
%% allowing the author to define a "short title" to be used in page headers.
\title{Visualizing Privacy-Relevant Data Flows in Android Applications}
%%
%% The "author" command and its associated commands are used to define
%% the authors and their affiliations.
%% Of note is the shared affiliation of the first two authors, and the
%% "authornote" and "authornotemark" commands
%% used to denote shared contribution to the research.
%% The abstract is a short summary of the work to be presented in the
%% article.

\author{Mugdha Khedkar}
\affiliation{%
  \institution{\textit{Heinz Nixdorf Institute \\ Paderborn University}}
  \city{Paderborn}
  \country{Germany}
}
\email{mugdha.khedkar@uni-paderborn.de}

\author{Michael Schlichtig}
\affiliation{%
  \institution{\textit{Heinz Nixdorf Institute \\ Paderborn University}}
  \city{Paderborn}
  \country{Germany}
}
\email{michael.schlichtig@uni-paderborn.de}

\author{Santhosh Mohan}
\affiliation{%
  \institution{\textit{Heinz Nixdorf Institute \\ Paderborn University}}
  \city{Paderborn}
  \country{Germany}
}
\email{santhosh.mohan@uni-paderborn.de}

\author{Eric Bodden}
\affiliation{%
  \institution{\textit{Heinz Nixdorf Institute \\ Paderborn University and Fraunhofer IEM}}
  \city{Paderborn}
  \country{Germany}
}
\email{eric.bodden@uni-paderborn.de}

\begin{abstract}
  Android applications collecting data from users must protect it according to the current legal frameworks. Such data protection has become even more important since in 2018 the European Union rolled out the General Data Protection Regulation (GDPR). Since app developers are not legal experts, they find it difficult to integrate privacy-aware practices into source code development. %Moreover, with the European Union's proposed Cyber Resilience Act on the horizon, app developers will soon face the challenge of writing code that complies with even more stringent security and privacy standards. 
  Despite these legal obligations, developers have limited tool support to reason about data protection throughout the development process.

  This paper explores the use of static program slicing and software visualization to analyze privacy-relevant data flows in Android apps.
  We introduce \emph{SliceViz}, a web tool that analyzes an Android app by slicing all privacy-relevant data sources detected in the source code on the back-end.
  It then helps developers by visualizing these privacy-relevant program slices.

  We conducted a user study with 12 participants demonstrating that SliceViz effectively aids developers in identifying privacy-relevant properties in Android apps.
  Our findings suggest that program slicing can be employed to reason about privacy-relevant data flows in Android apps.
  With further usability improvements, developers can be better equipped to handle privacy-sensitive information.
\end{abstract}

%%
%% The code below is generated by the tool at http://dl.acm.org/ccs.cfm.
%% Please copy and paste the code instead of the example below.
%%

%%
%% Keywords. The author(s) should pick words that accurately describe
%% the work being presented. Separate the keywords with commas.
%% A "teaser" image appears between the author and affiliation
%% information and the body of the document, and typically spans the
%% page.

%\received{20 February 2007}
%\received[revised]{12 March 2009}
%\received[accepted]{5 June 2009}

%%
%% This command processes the author and affiliation and title
%% information and builds the first part of the formatted document.
\maketitle

\input{Sections/1_Introduction}
\input{Sections/2_Problem}

\input{Sections/4_Approach}

\input{Sections/5_Implementation}

\input{Sections/6_Evaluation}

\input{Sections/3_RelatedWork}
\input{Sections/7_Limitations}

\input{Sections/8_Conclusion}

%%
%% The acknowledgments section is defined using the "acks" environment
%% (and NOT an unnumbered section). This ensures the proper
%% identification of the section in the article metadata, and the
%% consistent spelling of the heading.

%%
%% The next two lines define the bibliography style to be used, and
%% the bibliography file.
\bibliographystyle{ACM-Reference-Format}
\bibliography{sample-base}

\end{document}

%% file: Sections/1_Introduction.tex
\section{Introduction}
%\mk{@Eric, page limit is 10. +2 pages for references}
%\ms{Add A-Mobile Paper as motivation?}
%\mk{Cited. Added explicitly in research problem}
Any software that reaches the European market has to adhere to the General Data Protection Regulation (GDPR)~\cite{gdpr}, known as the \textit{toughest privacy and security law in the world}~\cite{whatisgdpr}. 
Moreover, with the EU's Cyber Resilience Act (CRA)~\cite{cra} on the horizon, software developers will soon face the challenge of writing code that complies with stricter security and privacy standards. 
These regulations extend to mobile applications that gather data from users in the EU. %According either piece of legislation, GDPR and CRA, data breaches in these apps can imply very severe fines.

The GDPR defines personal data as \textit{any information relating to an identified or identifiable natural person, a data subject}, %where an identifiable natural person is one who can be identified, directly or indirectly, by reference to an identifier such as a name, an identification number, etc. 
and imposes obligations on the access, storage and processing of such data. %\ms{There could be a cleaner transition between these two section. As I understand, CRA adds more tasks to devfelopers besides GDPR.}
Under the GDPR, data protection violations can result in severe financial penalties~\cite{penalties}. 
If these vulnerabilities 
%\eb{do not usually rather vulnerabilities cause violations?}
and data leaks cause volations, additional fines may be levied under the CRA. 
Companies thus must identify and secure privacy-critical %\eb{privacy-critical?} 
data-processing code, and prioritize testing efforts accordingly. 
%The main challenge is identifying this sensitive code early in the development lifecycle. %\eb{These arguments are motivational, should come at the top, not here.}
%\eb{Is that so? Can we quote somebody on this?}

The growing demand for privacy by design~\cite{pbd}, both by end users and by the GDPR, necessitates that app developers %\ms{For me it is not clear here, why we are already limited to app developers. Maybe mentioning DSS earlier or staying more general in the beginning of the inroduction makes sense.} 
use state-of-the-art technical measures to protect their users’ privacy. 
The legal description of the GDPR is very complex and lengthy and hence it can be difficult for app developers to understand. 
Since they lack legal expertise, app developers may be left wondering which technical measures need to be taken for which categories of user input data. 
In Android, merely including data-hungry third-party libraries may result in inadvertent data sharing~\cite{6979855}, highlighting the importance of careful consideration in app development. 

Advocating for transparency, Article §13~\cite{art13} of the GDPR mandates that the collection and processing of personal data be disclosed to the user through documents such as privacy policies. Therefore, all Android apps must provide a privacy policy which explains how apps collect, share, and process user data. However, these privacy policies are often long and vague, and may not always be authored by the app developers. Several studies~\cite{privacypolicytrust,automatedriskanalysis,guileak,ppviolationappcode,PTPDroid} have consistently shown significant discrepancies between privacy policies and the actual source code, undermining their accuracy and misleading users.

%Data Safety Section
To address these inaccuracies, Google introduced the data safety section (DSS)~\cite{data}, which shifts the responsibility of privacy-relevant reporting to app developers. 
This necessitates the completion of a form, outlining how apps collect, share, and secure user data. 
App developers also need to share if the app uses data protection measures such as encryption or on-demand data deletion. 
Without completing this form, an app cannot be listed on the Play Store. 
However, the process is manual and often error-prone~\cite{datalabels}, creating inconsistencies between what is reported and how privacy-relevant data is actually handled, which in turn gives users a false sense of privacy. 
To close this gap, developers require technical support to understand and accurately report the data practices of their apps. 
\begin{comment}
The form provided by Google, essential for adhering to the data safety section, consists of three main sections: \textit{data sharing, data collection, and security practices} (cf.~Figure~\ref{fig:dss_1}) and is a concise representation of privacy-relevant information. 
Within this form, user data is classified into different \textit{data categories} such as location and personal information%\ms{You could name the examples from Figure 1 here.} \mk{Like this?}
, which are further divided into specific \textit{data types} such as approximate location and user IDs (cf.~Figure~\ref{fig:dss_2}). 
This data can be collected or shared for different \textit{purposes} (chosen from a fixed set).

\begin{figure}[t]
    \centering
    \begin{subfigure}[b]{0.22\textwidth}
         \centering
         \includegraphics[width=\textwidth]{Figures/DS_0_annotated.jpg}
         \caption{Data safety section}
         \label{fig:dss_1}
     \end{subfigure}
     \hspace{0.2cm}
     \begin{subfigure}[b]{0.22\textwidth}
         \centering
         \includegraphics[width=\textwidth]{Figures/DS_1_annotated.jpg}
         \caption{Classification of data}
         \label{fig:dss_2}
     \end{subfigure}
    \centering
    \caption{An example of the data safety section of an Android app.}%\ms{The caption does not directly make sense with the text reference}}
    \label{fig:dss}
\end{figure}
\end{comment}

In this paper, we present an approach that uses static analysis to identify and evaluate privacy-relevant data in Android applications. 
Static analysis checks the source code thoroughly before execution, covering all possible execution paths, offering the potential to eventually provide legally useful guarantees. 

Since its introduction in the 1980s~\cite{Weiser, Weiser1}, the static analysis technique of program slicing has been used to assist developers in tasks such as debugging, testing, finding bugs, and understanding the code. 
Slicing systematically identifies a subset of the program that is relevant to a specific program location. 
Our approach applies forward static slicing to trace how privacy-relevant data propagates through the codebase. 
Specifically, it identifies statements and methods that are affected by sources of privacy-sensitive data, revealing dependencies and control flows that may impact privacy compliance.
%identifies and highlights all the statements that could be affected by a specific program location \cite{thinslicing}. \eb{This is Forward Slicing, in particular.}
%Our approach explores the use of forward static program slicing to identify parts of the source code affected by sources of privacy-relevant data. 
Our hypothesis is that visualizing the resulting program slices can help developers understand how privacy-relevant data flows through the source code and determine which privacy-relevant methods are reachable along the way. 
This, for instance, will allow developers to choose secure measures to protect privacy-relevant data. 
Although this analysis may appear similar to taint analysis at first glance, our approach differs in fundamental ways—particularly in how it handles sources and sinks. 
We elaborate on these distinctions in \Cref{relatedwork}.
%\eb{ (*) see below}

To implement this, we introduce SliceViz, a web-based tool that analyzes Android apps by slicing all privacy-relevant data sources detected in the source code. 
It then assists developers by visualizing these privacy-relevant program slices in both Jimple (a Java intermediate representation~\cite{soot}), and Java. 
Multiple studies~\cite{khedkarmobilesoft,avproposal} have emphasized the need for privacy-related tools that are accessible not only to developers but also to other stakeholders such as Data Protection Officers (DPOs), auditors, and legal experts. 
SliceViz was conceived as part of a broader project~\cite{assessorview} designed to support both technical and non-technical audiences. 
However, in this paper we focus specifically on the developer perspective. 
Developers are the first point of responsibility under DSS, and supporting them lays the groundwork for adapting SliceViz’s outputs to non-technical stakeholders in the broader project. 
To accommodate both groups, SliceViz operates directly on Android APKs—removing the requirement for local source code—and generates results that can be tailored for different audiences. 

In this work, we conduct a user study with 12 participants to evaluate how effectively SliceViz helps developers understand data protection in Android apps. 
The study shows that SliceViz is effective at helping developers identify and reason about privacy-relevant properties. 
Additionally, developers feel better supported by the Java program slices than Jimple, as they are more concise and align with the source code. 

To summarize, this work makes the following contributions:
\begin{itemize}
    \item We present SliceViz, a web tool that statically slices an Android app at the back-end, and assists developers by visualizing the privacy-relevant program slices.
  %  \item We evaluate PriBaSE and SliceViz on 36 applications, analyzing whether the resultance program slices provide insights into privacy-relevant data collection and logging.
    \item With a user study on 12 participants, we show that SliceViz aids developers in identifying and observing privacy-relevant properties in Android apps. %two primary challenges faced by developers.
 %   \item We introduce an optimized version of PriBaSE to help SliceViz reduce visual load during slice visualization.
  %  \item We implement usability enhancements to address two main issues reported by users: limited contextual information, and the complexity of the Jimple view. 
\end{itemize}

Artifacts are available at \href{https://doi.org/10.5281/zenodo.15487436}{https://doi.org/10.5281/zenodo.15487436}.

The remainder of the paper is organized as follows: %Section~\ref{motivation} discusses a motivating example. 
In \Cref{problem}, we use a motivating example to introduce the problem. %In Section~\ref{privacyrelevantdata}, we explain our definition of privacy-relevant data. 
We discuss our approach in \Cref{approach}. 
We discuss the implementation details in \Cref{impl}, and evaluate our approach in \Cref{eval}. 
We describe the related work in \Cref{relatedwork}. 
%In Section~\ref{usability} we explain our usability improvements. 
We explain the limitations of our approach in \Cref{limitations}, and conclude in \Cref{conclusion}.

%% file: Sections/2_Problem.tex
\section{Problem Statement}%\ms{Shouldn't that be the problem with example as subsection? OR is this because of Onward?}
\label{problem}

We next use a motivating example to explain our problem statement. 

\subsection{Motivating Example}
\label{example}

Alice, a Java programmer unfamiliar with legal privacy frameworks, and Bob, a
%\eb{I think that even a more experienced DPO would face the challenges that you describe next}
 Data Protection Officer (DPO), are tasked with ensuring that their company’s Android app complies with the General Data Protection Regulation (GDPR). 
As part of this effort, Bob must conduct a Data Protection Impact Assessment (DPIA)~\cite{dpia}, as mandated under Article §35 of the GDPR~\cite{art35}. 
A DPIA requires systematically answering questions about how the app collects, processes, stores, shares, deletes, and pseudonymizes personal data.

However, Alice and Bob face a fundamental challenge: the technical view of the app’s codebase does not easily translate into the legal concepts required for the DPIA. 
Alice may know where and how the app handles data, but she struggles to determine which flows are considered privacy-relevant under GDPR. 
Bob, on the other hand, understands the legal obligations but lacks the technical visibility needed to verify whether the app’s implementation complies with the GDPR.

Without dedicated tool support, they must rely on time-consuming, manual investigations of the code and extensive back-and-forth discussions to align technical details with legal requirements. 
This process is error-prone, hard to scale, and burdensome for smaller organizations without established compliance workflows.

What Alice and Bob wish they had is a way to automatically analyze the app, highlight privacy-relevant data flows, and generate outputs that both developers and non-technical stakeholders can use to decide whether additional compliance steps—such as a new DPIA—are necessary.

\subsection{Evidence of a Widespread Tooling Gap}
\label{evidence}

As Alice and Bob’s experience illustrates, bridging the gap between technical software artifacts and legal privacy compliance is difficult. %, especially when only an APK is available. 
Source code may not be easily accessible in many organizations—either because audits are performed by external Data Protection Officers (DPOs), or because legacy development practices make it hard to track the exact code deployed. 
A legal expert we interviewed further emphasized that this constraint is not only practical but also legally relevant: \textit{``The GDPR and national regulations in other countries have a lot of provisions in their data protection rules, protecting IP protected content. 
And the fact that you don't need to have the source, the main source, it's really interesting as to protect IP protected content.''} 
Without source code access, developers and DPOs must rely on manual inspection, incomplete documentation, or lengthy discussions to identify how personal data is collected, processed, and shared.

%bridging the gap between technical code understanding and legal privacy compliance is difficult, often requiring manual, error-prone collaboration between developers and legal experts. 
Our observations are not isolated. 
As part of a separate study for the broader project, we conducted interviews with 16 Data Protection Officers (DPOs) and legal experts, which confirmed that such challenges are widespread and systemic. 
One DPO emphasized: \textit{``Ideally, developers and legal teams should collaborate and conduct a DPIA before starting processing activities. But as my professional experience is concerned, that is the biggest void in the market. There is no communication between developers and lawyers—they are in two completely different worlds. Privacy by design requires developers to consider legal principles, but this rarely happens.''} 
Another DPO underscored the absence of automated tools for this collaboration, stating: \textit{``I have never seen automated code scanning from a privacy point of view (so far). Security tools detect some privacy issues but they are more focused on security, not on GDPR compliance part apart from Article §32~\cite{art32} (security of processing).''} 
Another DPO added, \textit{``I remember looking at tools that would look through code and then give you certain information but nothing specifically tailored to DPIAs.
I remember some early prototypes (2020, maybe), but nothing that ever got into production.''}

This lack of tool support also frustrates developers. 
In a survey of several dozen Android app developers we conducted, one developer remarked: \textit{``Many developers lack tools to automatically track and audit data usage, making it challenging to identify all collected or shared data.''} 
A recent empirical study~\cite{buggingyousomuch} demonstrated that software developers often fail to fulfill privacy requirements, and rarely contact privacy experts for assistance. 
Franke et al.~\cite{franke2024} surveyed 56 open-source developers to understand their experiences of GDPR. 
Their observations highlight the need for tools that support GDPR implementation and compliance in open-source software. 

Together, these findings highlight a systemic tooling gap: developers and DPOs lack solutions that can statically analyze software and produce privacy-relevant insights. % without requiring source code. 
Closing this gap could substantially reduce the manual, error-prone effort currently needed for tasks such as DPIAs and external audits.

\subsection{Research Question}
\label{rq}

Through this work, we aim to answer the following research question: 

\textit{How effectively does the combination of static program slicing and source code visualization help developers understand data protection in Android apps?} %Are current methods sufficient, or should we design new static analysis techniques?

Static program slicing can isolate smaller, privacy-relevant sections of the code that involve personal data collection, processing, and sharing. 
It is thus plausible to assume that visualizing these slices can help developers navigate large codebases more efficiently, improving their understanding of data practices and facilitating communication with legal teams. 
%Alice had to randomly sample the source code, since it was too large to be examined manually. 
%Access to smaller, more manageable privacy-relevant program slices could now potentially enhance her understanding of the source code. 
%We will investigate how visualizing these slices can facilitate collaboration between Alice and Bob for simplifying DPIAs. 

%% file: Sections/4_Approach.tex
\section{Approach}
\label{approach}

\begin{figure*}[t]
    \begin{center}
        \includegraphics[width=0.7\textwidth]{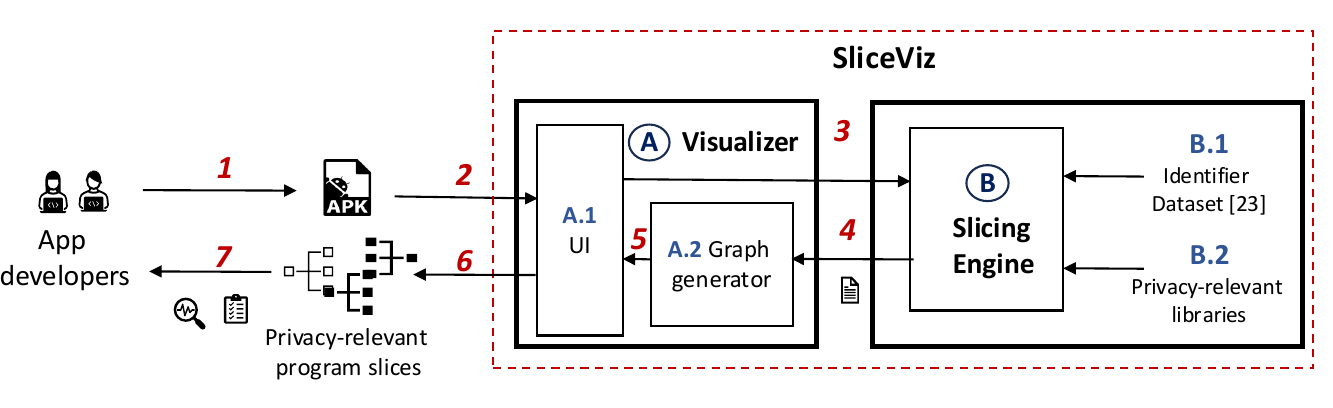}
        \caption{Overview of our approach and its components.}
        \label{fig:workflow}
    \end{center}
\end{figure*}

\begin{figure}[t]
    \begin{center}
        \includegraphics[width=0.52\textwidth]{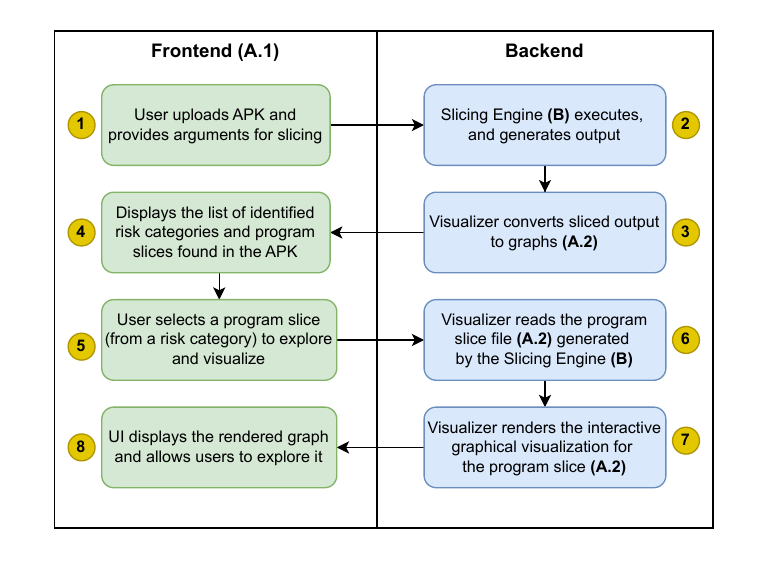}
        \caption{Workflow of SliceViz.}%\ms{Backend is A.2 and B? Maybe add this; Frontend is A.1}
        \label{fig:integratedworkflow}
    \end{center}
\end{figure}

Our approach to achieve the goals discussed in \Cref{problem} comprises two key components: \textbf{Visualizer} (Module\textcolor[HTML]{4472C4}{\textbf{~\textcircled{A}}} in \Cref{fig:workflow}), and \textbf{Slicing Engine} (Module\textcolor[HTML]{4472C4}{\textbf{~\textcircled{B}}} in \Cref{fig:workflow}).
SliceViz analyzes APKs, and not local source code, since it is a part of a broader project to support diverse stakeholders—such as auditors and external DPOs—who lack access to source code but still require tools for assessing privacy compliance.

Initially, when users upload an APK in SliceViz, Visualizer's user interface (\textcolor[HTML]{4472C4}{\textbf{A.1}}) processes user requests and executes the Slicing Engine (\textcolor[HTML]{4472C4}{\textbf{\textcircled{B}}}) with selected command-line options, such as specific risk sources, control dependencies, timeouts, and maximum node limits for slices.
It then parses the Slicing Engine's output, and visualizes the control and data dependencies within the sliced source code in Jimple (a Java intermediate representation~\cite{soot}) and Java.

\subsection{Visualizer}
\label{sliceviz}

Visualizer (\textcolor[HTML]{4472C4}{\textbf{\textcircled{A}}} in \Cref{fig:workflow}) is the entry point for users, allowing them to upload APKs.
The user interface (\textcolor[HTML]{4472C4}{\textbf{A.1}}) manages user inputs and interacts with the Slicing Engine (\textcolor[HTML]{4472C4}{\textbf{\textcircled{B}}}) by executing the analysis based on selected parameters.

%and displays a summary of the results on its dashboard. 

\begin{figure}[t]
    \begin{center}
        \includegraphics[width=0.5\textwidth]{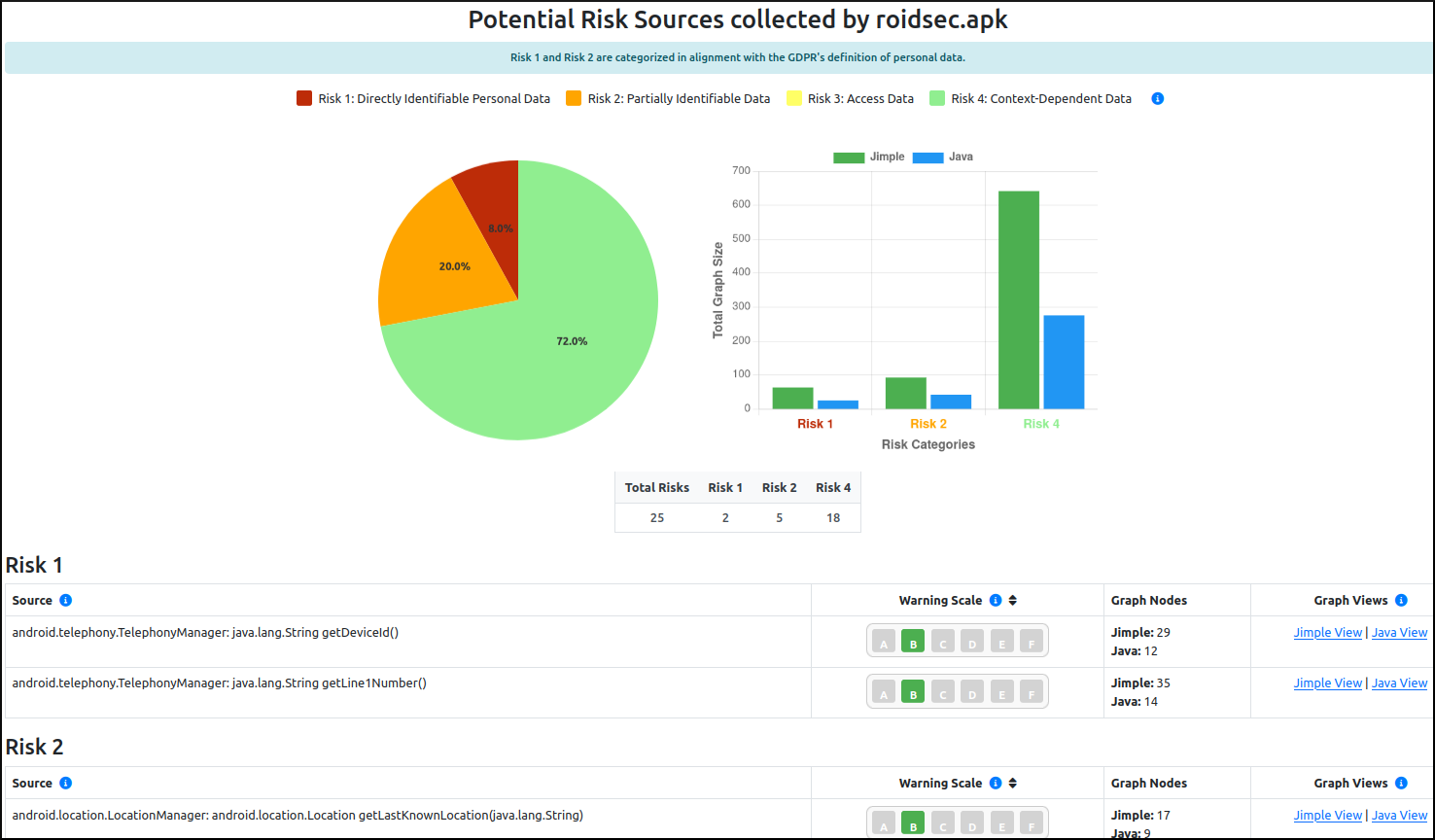}
        \caption{SliceViz \textit{interactive dashboard} for Roidsec~\cite{taintbench}.}
        \label{fig:slicevizdashboard}
    \end{center}
\end{figure}

\begin{table*}[!ht]
    \centering
    \small
    %\begin{tabular}{c@{\hspace{5pt}}|l@{\hspace{5pt}}|l@{\hspace{5pt}}|m{5cm}@{\hspace{5pt}}|m{7cm}@{\hspace{5pt}}}

    \caption{Warning scale levels displayed on the dashboard for every slice, and its legal privacy implications.}
    \label{tab:WarningScale}
    \begin{tabular}{cllm{5cm}m{7cm}}

        \toprule
        \textbf{Level}               & \textbf{Color} & \textbf{Risk}            & \textbf{Slice Property}                                                                                                   & \textbf{Legal privacy implication(s)}                                                                                                                                  \\ \midrule
        \cellcolor[HTML]{4CAF50}{A}  & Green          & Very low privacy risk    & Data collection, but no processing operations.                                                                            & Possibility of data minimization (GDPR Article §4~\cite{art4}).                                                                                                        \\ \midrule
        \cellcolor[HTML]{4CAF50}{B}  & Green          & Low privacy risk         & \begin{tabular}[c]{@{}l@{}} Data collection and string manipulations, \\but no other processing operations. \end{tabular} & Possibility of data minimization (GDPR Article §4~\cite{art4}).                                                                                                        \\ \midrule
        \cellcolor[HTML]{FFD700}{C}  & Yellow         & Moderate privacy risk    & At least one data storage or processing operation.                                                                        & \begin{tabular}[c]{@{}l@{}} Ensure data protection according to GDPR Article §25~\cite{art25}.\\ Document data usage (GDPR Article §13~\cite{art13}). \end{tabular}    \\ \midrule
        \cellcolor[HTML]{FFD700}{D}  & Yellow         & Significant privacy risk & Multiple data storage or processing operations.                                                                           & \begin{tabular}[c]{@{}l@{}} Ensure data protection according to GDPR Article §25~\cite{art25}. \\ Document data usage (GDPR Article §13~\cite{art13}). \end{tabular}   \\ \midrule
        \cellcolor[HTML]{FF6347} {E} & Red            & High privacy risk        & Data shared with third-party APIs at least once.                                                                          & \begin{tabular}[c]{@{}l@{}} Ensure data protection according to GDPR Article §25~\cite{art25}. \\Document data sharing (GDPR Article §13~\cite{art13}). \end{tabular}  \\ \midrule
        \cellcolor[HTML]{FF6347} {F} & Red            & Very high privacy risk   & Data shared with third-party APIs multiple times.                                                                         & \begin{tabular}[c]{@{}l@{}} Ensure data protection according to GDPR Article §25~\cite{art25}. \\Document data sharing (GDPR Article §13~\cite{art13}).  \end{tabular} \\ \bottomrule
    \end{tabular}
\end{table*}

Upon completion of the slicing process, Visualizer consumes the report in form of text files generated by the Slicing Engine, and displays a summary of the results on SliceViz's dashboard (cf.~\Cref{fig:slicevizdashboard}).
Although graph rendering is central to visualizing program slices, equally important is the SliceViz dashboard, which consolidates privacy-related practices—such as data collection sources, processing operations, and associated risks.
It provides developers with a high-level overview that supports faster comprehension and informed decision-making.
Each identified program slice is assigned two distinct risk ratings.

First, it is classified based on the risk of its \textbf{source}, according to the risk categories defined in the Identifier Dataset (\textcolor[HTML]{4472C4}{\textbf{B.1}}).
We explain this classification in \Cref{subsec_PrivacyBasedSlicingEngine}.
Second, it is assigned a warning level using a six-tier Warning Scale introduced by us (A–F, cf.~\Cref{tab:WarningScale}) %\ms{we introduced this scale, right? Then this should be mentioned and we should provide the reasoning, e.g., in the artifact. From Table 1 a reader cannot retrieve the exact definition. Further, providing an example could help a reader'S understanding.} 
based on the \textbf{processing operations} performed on the privacy-relevant data, which helps prioritize actions.
Levels A and B denote low risks, levels C and D represent moderate risks, while levels E and F indicate high risks requiring immediate attention.
Specific messages accompanying each level provide detailed insights into the potential risks.
For example, since the Identifier Dataset categorizes any method collecting \textit{device ID} as a \colorbox{lightred!40}{risk 1} source, any program slice originating from a \textit{device ID} source will be categorized as \colorbox{lightred!40}{risk 1}.
However, within this category, the warning level (A–F) may vary depending on the processing operations in the program slice (cf.~\Cref{tab:WarningScale}).

Users can then select a program slice to view its graphical visualization in SliceViz (workflow shown in \Cref{fig:integratedworkflow}).
%\eb{Fig 2 says that users select a risk category, not a risk source. That's not the same, is it?}
%The visualized graph allows users to check documentation of the data source and offers other documentation explaining the risk-based classification of data.
% (cf.~Figure~\ref{fig:roidsecslice}). 
%SliceViz also offers \textit{customization options}, allowing users to exclude control dependencies and generate data flow graphs for focused code analysis (Figure~\ref{fig:slicevizhome}). 

%\eb{The timeout is really in terms of seconds? Or is this a max length of a slice?}

\subsection{Slicing Engine}
\label{subsec_PrivacyBasedSlicingEngine}

% \santhosh{I added this based on our disc. but seems repetetive to \Cref{subsec_PrivacyBasedSlicingEngine}, please confirm}

In SliceViz, the Slicing Engine (\textcolor[HTML]{4472C4}{\textbf{\textcircled{B}}} in \Cref{fig:workflow}) functions as the analysis core, performing static examination of APK files and producing their corresponding dependency graphs.
% Slicing Engine (\textcolor[HTML]{4472C4}{\textbf{\textcircled{B}}} in \Cref{fig:workflow}) is the back-end of SliceViz, responsible for performing static analysis on APKs and generating dependence graphs. 
It extends Jicer \cite{jicer}, an open-source static program slicer for Android that analyzes Jimple, an intermediate representation of Java code \cite{soot}. The extension introduces two key components: an \emph{Identifier Dataset} (\textcolor[HTML]{4472C4}{\textbf{B.1}}) to match system API calls with privacy labels, and a collection of \emph{privacy-relevant libraries} package names to identify privacy-relevant methods (\textcolor[HTML]{4472C4}{\textbf{B.2}}).

% Slicing Engine (\textcolor[HTML]{4472C4}{\textbf{\textcircled{B}}} in \Cref{fig:workflow}) is the back-end of SliceViz, responsible for performing static analysis on APKs and generating dependence graphs.
% It requires two elements, namely an \emph{Identifier Dataset} (\textcolor[HTML]{4472C4}{\textbf{B.1}}) to match system API calls with privacy labels, and a collection of \emph{privacy-relevant libraries} package names to identify privacy-relevant methods (\textcolor[HTML]{4472C4}{\textbf{B.2}}).

%\eb{what exactly do you mean by a collection? do you need the bytecode implementation of those libraries? or their class names? or what else?} \mk{Package names}
The Slicing Engine employs both \textcolor[HTML]{4472C4}{\textbf{B.1}} and \textcolor[HTML]{4472C4}{\textbf{B.2}} to identify privacy-relevant sources, and then slices the Android app from these privacy-relevant sources. %\eb{ends a bit abruptly...}. 
It is implemented as Java code and seeks to answer two primary questions that developers might have:
\begin{enumerate}
    \item Which data collected by the app should be considered and reported as high-risk?
    \item How does this data flow through the source code, and does it flow into data processing methods (eg. analytics, advertisement, authorization methods)?
\end{enumerate}

%\stilltodo{MOBILSoft review: Library Collection: The paper lacks a clear methodology for collecting privacy-relevant APIs from third-party libraries. While it mentions using AppBrain statistics and related literature to select commonly used libraries, it fails to address the fundamental question of how privacy-relevant APIs are identified and collected within these libraries. The criteria for determining privacy relevance and mechanisms for ensuring API collection completeness remain undefined. This becomes particularly concerning in the context of code obfuscation, which the paper acknowledges as a limitation. The paper would benefit from a more thorough discussion of these challenges and potential approaches to addressing them.}
%\mk{I have no idea what to do about this. We are not talking about libraries in the identifier dataset since we talk about it in the PRICE paper.}

%Before we start the analysis, in order to label sources as potential sources of privacy-relevant data we require a dataset that can be used as a classification criterion. %We constructed two different datasets: an \textit{identifier keywords %\eb{see comment above} \mk{we named this dataset identifier keywords dataset since it has keywords which could be identifiers. The other dataset has API calls which could be identifiers. Should I change it?} dataset} for labeling UI data elements in UI analysis and an 
\textcolor[HTML]{4472C4}{\textbf{B.1}} \textbf{Identifier Dataset.} To answer the first question, the Slicing Engine uses our manually annotated \textit{Identifier Dataset}~\cite{asej_price}, which labels data sources based on a risk-based definition of privacy-relevant data.
%This dataset uses a risk-based definition of privacy-relevant data and labels system API calls in the app code. 
Personal data that can identify an individual or a device without the use of any additional information is categorized as \colorbox{lightred!40}{risk 1}, while data requiring additional information for identification is classified as \colorbox{lightorange!40}{risk 2}.
The dataset categories align with Google’s data safety section to simplify form completion.
%The data categories in the dataset match the categories in Google’s data safety section to simplify completion of the data safety form. 

\textcolor[HTML]{4472C4}{\textbf{B.2}} \textbf{Privacy-Relevant Libraries.} The Slicing Engine statically matches the system API calls against a list of packages of privacy-relevant third-party libraries and APIs.
Note that the Slicing Engine is designed as a proactive privacy by design~\cite{pbd} measure, integrating into the app developer's workflow to analyze privacy-relevant third party libraries and API calls early in the development process.
It operates before name mangling or obfuscation~\cite{obfuscationuse, niroshan2025empiricalstudycodeobfuscation}, which are typically applied later when packaging the app for deployment.

We consulted the AppBrain statistics~\cite{appbrain} and related literature~\cite{analyticslibrarysource,mudflow,tpl} to select the most commonly used libraries for pseudonymization, analytics, advertisements, authentication, network, input/output, email, image processing, data serialization and location. %Additionally, we select the most commonly used  libraries and 
According to GDPR, robust pseudonymization techniques involve creating pseudonymns that cannot be easily re-identified and reproduced by third parties~\cite{enisa2}.
Hashing, considered a weak technique for pseudonymization, can be reversed with a limited dictionary~\cite{hashingpitfalls}.
Grading pseudonymization functions based on these criteria can help app developers identify areas requiring additional data protection measures. %Table~\ref{tab:tpls} gives some examples of these libraries.

After the slicing process is complete, the Slicing Engine generates text files with privacy-relevant program slices and transfers them to Visualizer for further analysis.

Instead of relying on high-effort random sampling, Alice (cf.~\Cref{example}) can use SliceViz to quickly obtain a high-level overview of all privacy risks in the Android application. She can then examine each individual slice in detail, tracing how data flows within the app, and explain these flows to Bob. As a result, Alice and Bob can complete the DPIA with significantly less manual labor.

%% file: Sections/5_Implementation.tex
\section{Implementation}
\label{impl}

We now discuss the implementation details of SliceViz (cf.~\Cref{fig:workflow}).
\subsection{Visualizer}
\label{sliceviz}

Users upload an APK in SliceViz, and Visualizer's user interface (\textcolor[HTML]{4472C4}{\textbf{A.1}}) processes user requests and executes the Slicing Engine (\textcolor[HTML]{4472C4}{\textbf{\textcircled{B}}}) with selected command-line options.
The user interface (\textcolor[HTML]{4472C4}{\textbf{A.1}}) uses D3.js~\cite{d3.js}, a JavaScript library for visualizing data. D3.js offers customization, scalability, flexibility and operates at a low level while creating complex and dynamic visualization~\cite{d3paper}.

Once the Slicing Engine completes its execution, Visualizer consumes the text files generated by the Slicing Engine (\textcolor[HTML]{4472C4}{\textbf{\textcircled{B}}} in \Cref{fig:workflow}).
It parses them and visualizes the program slices as graphs, rendering key components such as metadata, nodes, edges, and other attributes on the screen (\textcolor[HTML]{4472C4}{\textbf{A.2}}).

\begin{figure*}[htbp]
    \centering
    \begin{subfigure}{0.8\linewidth}
        \centering
        \includegraphics[width=0.8\linewidth]{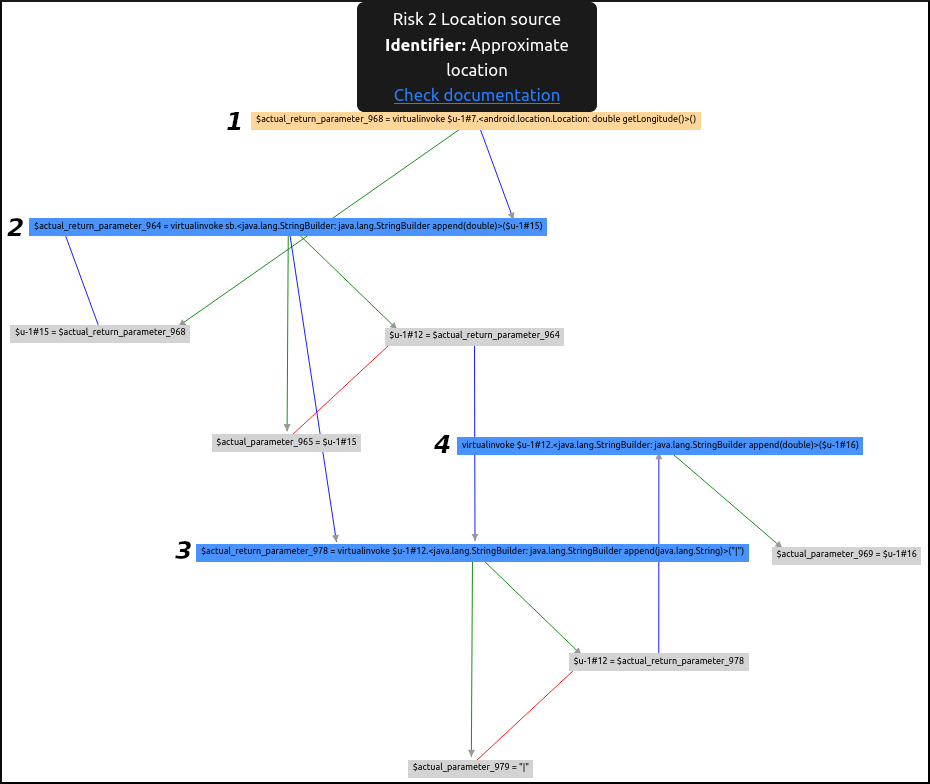}
        \caption{\textit{Jimple View} of a Roidsec slice.}
        \label{fig:first}
    \end{subfigure}

    \vspace{1em} % optional space between figures

    \begin{subfigure}{0.8\linewidth}
        \centering
        \includegraphics[width=0.8\linewidth]{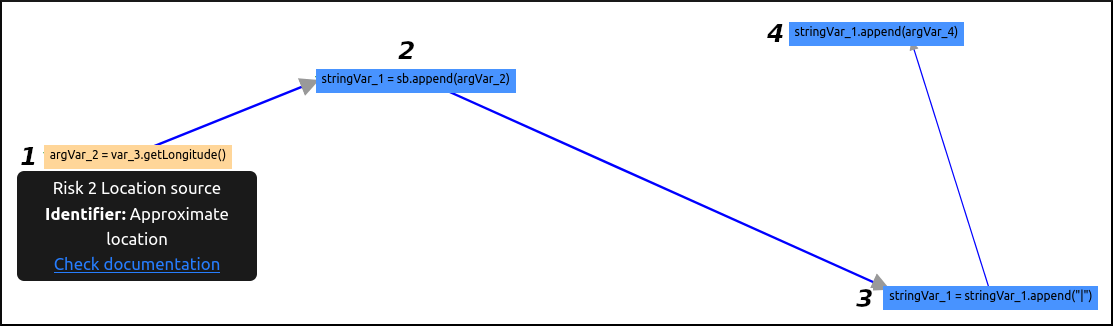}
        \caption{\textit{Java View} of a Roidsec slice.}
        \label{fig:second}
    \end{subfigure}
    \caption{SliceViz presenting the same program slice in both Jimple and Java Views. Nodes: source code statements, edges: control and data dependencies. Pop-ups reveal details for each node (pop-up for node 1 shown in the figure). Node 1 collects longitude (\colorbox{lightorange!40}{risk 2}), which is appended to a string (node 2). Nodes 3 and 4 append other variables. Unnumbered nodes in \Cref{fig:first} are spurious nodes specific to Jimple. The dashboard assigns a low-risk (\colorbox[HTML]{4CAF50}{B}) warning since only string manipulations occur, suggesting potential for data minimization (GDPR Article §4~\cite{art4}).}
    \label{fig:combined}
\end{figure*}

Users can then select a program slice to view its graphical visualization in SliceViz (workflow shown in \Cref{fig:integratedworkflow}).
%\eb{Fig 2 says that users select a risk category, not a risk source. That's not the same, is it?}
The visualized graph allows users to check documentation of the data source and offers other documentation explaining the risk-based classification of data.
\Cref{fig:combined} illustrates how SliceViz presents program slices in both Java and Jimple views.
At this stage, the figure is introduced as an integrated example; each view (cf.~\Cref{fig:first} and \Cref{fig:second}) will be discussed in more detail in \Cref{subsec_PrivacyBasedSlicingEngine}.
For brevity, we illustrate a simplified program slice example in this paper (cf.~\Cref{fig:combined}); more complex cases can be found in the accompanying artifact\footnote{\href{https://doi.org/10.5281/zenodo.15487436}{https://doi.org/10.5281/zenodo.15487436}}.
\begin{comment}
\begin{figure*}[t]
    \begin{center}
        \includegraphics[width=0.8\textwidth]{Figures/complexJavaView.png}
        \caption{\textit{Java View} of a Roidsec slice. Nodes: source code statements, edges: control and data dependencies. Pop-ups reveal details for each node (pop-up for node 1 shown in the figure). Node 1 collects the network operator name (\colorbox{lightorange!40}{risk 2}), which is copied (node 2), and appended to a string (node 3). Nodes 4–9 append other variables. The dashboard assigns a low-risk (\colorbox[HTML]{4CAF50}{B}) warning since only string manipulations occur, suggesting potential for data minimization (GDPR Article §4~\cite{art4}).}
        \label{fig:javaview}
    \end{center}
\end{figure*}
\end{comment}

\begin{comment}
\begin{figure*}[t]
    \begin{center}
        \includegraphics[width=0.8\textwidth]{Figures/RoidsecJimple.png}
        \caption{}
        \label{fig:javaview}
    \end{center}
\end{figure*}

\begin{figure*}[t]
    \begin{center}
        \includegraphics[width=0.8\textwidth]{Figures/RoidsecJava.png}
        \caption{Nodes: source code statements, edges: control and data dependencies. Pop-ups reveal details for each node (pop-up for node 1 shown in the figure). Node 1 collects longitude (\colorbox{lightorange!40}{risk 2}), which is appended to a string (node 2). Nodes 3 and 4 append other variables. The dashboard assigns a low-risk (\colorbox[HTML]{4CAF50}{B}) warning since only string manipulations occur, suggesting potential for data minimization (GDPR Article §4~\cite{art4}).}
        \label{fig:javaview}
    \end{center}
\end{figure*}
\end{comment}

\textbf{Early experiments -- SliceViz.}
The first implementation of SliceViz composed only of a simple UI, the Slicing Engine (\textcolor[HTML]{4472C4}{\textbf{\textcircled{B}}}) and the Jimple View.
For an initial assessment of the viability of our approach we conducted two pilot studies (involving 10 and 6 participants, respectively) to evaluate the usability of SliceViz.
While participants found the visualized program slices to be more efficient than manual code inspection, those unfamiliar with Jimple struggled to interpret the graphs generated by SliceViz.
Additionally, visualizations with more than 50 nodes became challenging to interpret, due to increased cognitive load, as proven by Yoghourdjian et al.~\cite{cognitiveload}.
Since Jimple breaks down each Java statement into multiple statements, the Jimple View resulted in larger slices.

To address this, we implemented an experimental \textit{Java View} that relies on the back-end to simplify and transform the data.
The Slicing Engine's transformers omit Jimple-specific nodes, reducing the size and complexity of the application dependence graph (ADG) while preserving essential privacy-relevant information.
Developers can now choose which view, \textit{Jimple} or \textit{Java}, they use.
Moreover, the UI provides three key \textit{customization options} to manage graph complexity and reduce visual overload:
\begin{enumerate}
    \item Toggle inclusion of control dependencies in the graph.
    \item Set a maximum analysis time limit for generating slices.
    \item Specify a maximum number of nodes allowed in a graph.
\end{enumerate}
%\eb{The timeout is really in terms of seconds? Or is this a max length of a slice?}

\subsection{Slicing Engine}
\label{subsec_PrivacyBasedSlicingEngine}

% \santhosh{Rephrase?} -- Rephrased in section 3.2, so keep as it is
Slicing Engine (\textcolor[HTML]{4472C4}{\textbf{\textcircled{B}}}) is the back-end of SliceViz, responsible for performing static analysis on APKs and generating dependence graphs.

The Slicing Engine is a command-line tool implemented in Java.
It is built on top of Jicer~\cite{jicer}, a static program slicer that works with Jimple, an intermediate representation of Java code.
Jicer is tailored to address Android-specific challenges, such as the absence of a main method, extensive use of callbacks, and inter-component communication~\cite{jicer}.
The Slicing Engine uses Jicer's functionality to statically construct an application dependence graph (ADG) from the APK, preserving control and data dependencies in the source code~\cite{slicingSDG}.
It introduces its own algorithm to match method signatures from the ADG with the \textit{Identifier Dataset} (\textcolor[HTML]{4472C4}{\textbf{B.1}}), labeling privacy-relevant data sources in the code.
Once labeled, the Slicing Engine uses Jicer’s slicing component to perform forward slicing on the ADG, starting from the identified privacy-relevant data sources.
This process yields forward slices representing the code components reachable from all labeled privacy-relevant data sources.
Finally, the Slicing Engine adds its own functionality to automatically label pseudonymization methods and other privacy-relevant methods within the sliced ADG using a privacy-relevant libraries dataset ({\textcolor[HTML]{4472C4}{\textbf{B.2}}).
It stores the sliced ADGs as text files, facilitating further analysis.

When the \textit{Java View} is enabled through the user interface, SliceViz triggers the Slicing Engine to execute custom transformations on the sliced ADGs.
These transformations remove intermediate Jimple-specific nodes, reduce graph size and complexity, and convert Jimple nodes into more readable, Java-like representations to improve program comprehension.
The transformed graphs present semantically correct Java-like statements (cf.~\Cref{fig:second}) instead of raw Jimple statements (cf.~\Cref{fig:first}), enabling developers to work effectively without needing prior Jimple knowledge.

Currently, the Slicing Engine transforms key Jimple statements into Java-like representations, including \emph{invoke}, \emph{assignment}, \emph{if}, \emph{return}, \emph{binary expressions}, \emph{invoke expressions}, and \emph{array variables}.
These statements are commonly involved in data flows and their processing, and they tend to be more complex to interpret in Jimple compared to Java.
Moreover, recent research~\cite{su2025codestatementsimplementprivacy} reveals that \emph{function calls}, \emph{expression statements}, \emph{if} and \emph{decl statements} implement privacy behaviors in Android apps.
We are actively working on expanding the mapping between Jimple and Java to improve the \textit{Java View}.
%\mk{This recent paper \href{https://arxiv.org/pdf/2503.02091}{https://arxiv.org/pdf/2503.02091} reveals that function calls, expression statements, if and decl statements are mostly relevant to privacy. Should we cite this paper?}
%\eb{Sure, makes sense}
%\eb{That is exactly what I would argue: These are exactly the kind of statements that are (1) commonly involved in data flows and their processing, and (2) that are more complex to understand in Jimple than they are in Java.}
%\ms{Do we have some data that those are the e.g., most frequent statements? OR, we contacted soot / soot\_up maintainers and they advised to start with these statements .... }
%\mk{We chose these statements because they occured the most in the slices we visualized and observed in the early experiments (where we visualized and manually analyzed more than 25 slices from 16 apps). Can we just say this, then? }
\begin{comment}
\begin{figure}[t]
    \begin{center}
        \includegraphics[width=0.47\textwidth]{Figures/JavaView.png}
        \caption{\textit{Java view} for the Roidsec Jimple slice.}
        \label{fig:javaview}
    \end{center}
\end{figure}
\end{comment}

%\mk{@EB, updated this as per our discussion.}

\textbf{Early experiments -- Slicing Engine.} At the initial assessment of SliceViz, we also tested the Slicing Engine on 36 applications: 5 apps from DroidBench~\cite{droidbench}, 25 apps from TaintBench~\cite{taintbench}, and 6 real-world Android apps available on Google Play Store.
While DroidBench contains hand-crafted, small micro benchmark apps, the TaintBench suite consists of real-world malware apps, which are similar with respect to size and features to apps downloadable from Google Play Store.
We observed that program slices were getting too large for manual evaluation.
Visualizing such large slices remains an open challenge, motivating the need for additional summarization techniques.
We hope to engage with the ICPC community to address this issue.
Consequently, for this paper, we did not extend the experiment to a larger set of real-world applications.

Instead, we explored modifications to the slicing process itself, specifically by comparing slice sizes generated with and without control dependencies in the ADG.
Excluding control dependencies led to a substantial reduction in slice sizes.
Based on this finding, we added a \textit{customization option} into the SliceViz user interface, enabling users to toggle the inclusion of control dependencies during slice construction.
In many cases, disabling control dependencies might be sensible, as control dependencies do not usually reveal private data itself but at best information about this private data.

\subsection{Correctness Check}

To ensure the correctness of SliceViz’s results, we first conducted sanity checks by manually verifying its output.
The experiments were performed on an Ubuntu 20.04.2 machine with an Intel(R) Core i7-10850H processor (6 cores).
In total, we evaluated SliceViz on 51 benchmark applications.

%\mk{@EB, added this additional ground truth check, where I manually looked into all slices to see if they match the program behavior of DroidBench apps. So now we have two checks: whether the detected source is, in fact, present in the app. And if the control flow and data flow detected is correct in the source code of DroidBench}

We first applied SliceViz to 18 DroidBench test cases~\cite{droidbench} and manually compared the generated program slices against the documented program behavior.
To ensure coverage, we selected at least one test case from each category of the microbenchmark suite.
No discrepancies were observed.
On average, SliceViz processed each test case in 34.38 seconds.

Next, we tested SliceViz on 33 applications from TaintBench~\cite{taintbench}, verifying the presence of detected risk sources against the ground truth.
We chose all apps from the TaintBench suite (33 out of 39) that had code size less than 100K lines.
This LOC threshold ensured that slices remained inspectable and comparable against the benchmark’s ground truth—something not feasible with larger real-world apps.
The sanity check revealed that all the sources detected by SliceViz matched with the sources present in the TaintBench ground truth.
On average, SliceViz processed each app in 41.85 seconds.

%% file: Sections/6_Evaluation.tex
\section{Evaluation}
\label{eval}

In Section~\ref{rq}, we defined our broad research question, which explores how program slicing and visualization can be combined to help developers understand data protection in Android apps.
In Sections~\ref{approach} and \ref{impl}, we introduced SliceViz, a web-based tool that operationalizes this combination by slicing privacy-related data sources and visualizing their flows in the app code.
To evaluate how this combination supports developers in practice, we conducted a user study that addresses the following research questions:

\textit{RQ1. To what extent does SliceViz help developers understand data protection in Android apps?}

This question evaluates SliceViz as a whole, capturing the combined effect of program slicing and visualization on developers’ ability to understand privacy-relevant aspects of source code.
By addressing this question, we assess how tooling that integrates slicing and visualization can support developers during the development and analysis of Android apps.

\textit{RQ2. To what extent does the experimental Java View better support developers in understanding the flow of privacy-relevant data within an app?}

RQ2 focuses on one central component of SliceViz—the Java View (cf.~Section~\ref{sliceviz})—which simplifies Jimple statements to improve program comprehension. By isolating this design choice, we examine how a Java-oriented visualization contributes to developers’ understanding of privacy-relevant data flows and helps explain the effects observed in RQ1.

In our evaluation, the baseline reflects the current practice: app developers performing these tasks manually, without tool support (cf.~\Cref{evidence}).
This involves manually inspecting source code to identify privacy-relevant data and assess processing activities—an approach that, as prior research~\cite{datalabels} and our motivating example (\Cref{example}) suggest, is time-consuming and may be error-prone.

\subsection{Experimental Setup}

To answer the research questions, we conducted a within-subjects user study~\cite{withinsubjects} with 12 participants.
The user study design was approved by the Ethics committee of our institution.
We initially attempted to recruit app developers who had participated in a survey we conducted as part of a separate study, but none responded within the available timeframe.
As a result, and given that students are considered reasonable proxies for software developers in usability studies~\cite{userstudy1,userstudy2,userstudy3}, we recruited Computer Science students instead.
The 12 participants of this user study are disjoint from the 16 participants of the pilot studies initially conducted for SliceViz.
Interested participants were first sent to a landing page informing them about the study’s purpose.
Then, they were shown our consent form where we informed them about how we would handle their data, their right to terminate the study at any time, and where they could give consent to participate in the study.
Finally, they could schedule select time slots for the user study.
Our consent form, privacy policy and data handling practices were approved by the Data Protection Office of our institution.

%All the participants were Computer Science students. 
At the beginning of the study, participants were given all the information about the study, and were asked to sign the \textit{informed consent} form.
They then answered questions on a Likert scale from 1 (Beginner) to 5 (Expert) about their experience (cf.~\Cref{tab:participants}).

\begin{table*}[t]
    \caption{Summary of participant experience. Filled circles indicate expertise from 1 (Beginner) to 5 (Expert).}
    \label{tab:participants}
    \centering
    \small
    %\begin{tabular}{m{1cm}m{1.5cm}|m{1.5cm}|m{1.5cm}|m{1.5cm}|m{1.5cm}|m{1.5cm}|m{1.5cm}m{1.5cm}}
    %\hline
    \begin{tabular}{ll@{\hspace{5pt}}|l@{\hspace{5pt}}|l@{\hspace{5pt}}|l@{\hspace{5pt}}|l@{\hspace{5pt}}|l@{\hspace{5pt}}|l}
        \toprule
        \textbf{Participant}                                                                                          & \textbf{\begin{tabular}[c]{@{}l@{}}Android app \\development \end{tabular}}                                   & \textbf{\begin{tabular}[c]{@{}l@{}}Java\\ development \end{tabular}} & \textbf{\begin{tabular}[c]{@{}l@{}}Jimple \\understanding \end{tabular}} & \textbf{\begin{tabular}[c]{@{}l@{}}Static analysis \\tools \end{tabular}} & \textbf{\begin{tabular}[c]{@{}l@{}} Graph \\ visualization \end{tabular}} & \textbf{\begin{tabular}[c]{@{}l@{}}Data privacy \end{tabular}} & \textbf{\begin{tabular}[c]{@{}l@{}}GDPR \end{tabular}} \\ \midrule % \textbf{\begin{tabular}[c]{@{}l@{}}Java dev\\ experience\\ (in years) \end{tabular}} 
        P01                                                                                                           & \tikzcircleblack{3pt} \tikzcircleblack{3pt} \tikzcirclewhite{3pt} \tikzcirclewhite{3pt} \tikzcirclewhite{3pt} &
        \tikzcircleblack{3pt} \tikzcircleblack{3pt} \tikzcircleblack{3pt} \tikzcirclewhite{3pt} \tikzcirclewhite{3pt} &
        \tikzcircleblack{3pt} \tikzcirclewhite{3pt} \tikzcirclewhite{3pt} \tikzcirclewhite{3pt} \tikzcirclewhite{3pt} &
        \tikzcircleblack{3pt} \tikzcirclewhite{3pt} \tikzcirclewhite{3pt} \tikzcirclewhite{3pt} \tikzcirclewhite{3pt} &
        \tikzcircleblack{3pt} \tikzcircleblack{3pt} \tikzcircleblack{3pt} \tikzcirclewhite{3pt} \tikzcirclewhite{3pt} &
        \tikzcircleblack{3pt} \tikzcircleblack{3pt} \tikzcircleblack{3pt} \tikzcirclewhite{3pt} \tikzcirclewhite{3pt} &
        \tikzcircleblack{3pt} \tikzcircleblack{3pt} \tikzcircleblack{3pt} \tikzcirclewhite{3pt} \tikzcirclewhite{3pt}                                                                                                                                                                                                                                                                                                                                                                                                                                                                                                                                                     \\

        P02                                                                                                           & \tikzcircleblack{3pt} \tikzcircleblack{3pt} \tikzcircleblack{3pt} \tikzcircleblack{3pt} \tikzcirclewhite{3pt} &
        \tikzcircleblack{3pt} \tikzcircleblack{3pt} \tikzcircleblack{3pt} \tikzcirclewhite{3pt} \tikzcirclewhite{3pt} &
        \tikzcircleblack{3pt} \tikzcirclewhite{3pt} \tikzcirclewhite{3pt} \tikzcirclewhite{3pt} \tikzcirclewhite{3pt} &
        \tikzcircleblack{3pt} \tikzcircleblack{3pt} \tikzcirclewhite{3pt} \tikzcirclewhite{3pt} \tikzcirclewhite{3pt} &
        \tikzcircleblack{3pt} \tikzcircleblack{3pt} \tikzcirclewhite{3pt} \tikzcirclewhite{3pt} \tikzcirclewhite{3pt} &
        \tikzcircleblack{3pt} \tikzcircleblack{3pt} \tikzcircleblack{3pt} \tikzcirclewhite{3pt} \tikzcirclewhite{3pt} &
        \tikzcircleblack{3pt} \tikzcirclewhite{3pt} \tikzcirclewhite{3pt} \tikzcirclewhite{3pt} \tikzcirclewhite{3pt}                                                                                                                                                                                                                                                                                                                                                                                                                                                                                                                                                     \\

        P03                                                                                                           & \tikzcircleblack{3pt} \tikzcircleblack{3pt} \tikzcircleblack{3pt} \tikzcirclewhite{3pt} \tikzcirclewhite{3pt} &
        \tikzcircleblack{3pt} \tikzcircleblack{3pt} \tikzcircleblack{3pt} \tikzcirclewhite{3pt} \tikzcirclewhite{3pt} &
        \tikzcircleblack{3pt} \tikzcirclewhite{3pt} \tikzcirclewhite{3pt} \tikzcirclewhite{3pt} \tikzcirclewhite{3pt} &
        \tikzcircleblack{3pt} \tikzcircleblack{3pt} \tikzcirclewhite{3pt} \tikzcirclewhite{3pt} \tikzcirclewhite{3pt} &
        \tikzcircleblack{3pt} \tikzcircleblack{3pt} \tikzcirclewhite{3pt} \tikzcirclewhite{3pt} \tikzcirclewhite{3pt} &
        \tikzcircleblack{3pt} \tikzcircleblack{3pt} \tikzcircleblack{3pt} \tikzcirclewhite{3pt} \tikzcirclewhite{3pt} &
        \tikzcircleblack{3pt} \tikzcircleblack{3pt} \tikzcircleblack{3pt} \tikzcirclewhite{3pt} \tikzcirclewhite{3pt}                                                                                                                                                                                                                                                                                                                                                                                                                                                                                                                                                     \\

        P04                                                                                                           & \tikzcircleblack{3pt} \tikzcirclewhite{3pt} \tikzcirclewhite{3pt} \tikzcirclewhite{3pt} \tikzcirclewhite{3pt} &
        \tikzcircleblack{3pt} \tikzcircleblack{3pt} \tikzcircleblack{3pt} \tikzcirclewhite{3pt} \tikzcirclewhite{3pt} &
        \tikzcircleblack{3pt} \tikzcircleblack{3pt} \tikzcircleblack{3pt} \tikzcirclewhite{3pt} \tikzcirclewhite{3pt} &
        \tikzcircleblack{3pt} \tikzcircleblack{3pt} \tikzcirclewhite{3pt} \tikzcirclewhite{3pt} \tikzcirclewhite{3pt} &
        \tikzcircleblack{3pt} \tikzcircleblack{3pt} \tikzcircleblack{3pt} \tikzcircleblack{3pt} \tikzcirclewhite{3pt} &
        \tikzcircleblack{3pt} \tikzcircleblack{3pt} \tikzcircleblack{3pt} \tikzcircleblack{3pt} \tikzcirclewhite{3pt} &
        \tikzcircleblack{3pt} \tikzcircleblack{3pt} \tikzcircleblack{3pt} \tikzcircleblack{3pt} \tikzcirclewhite{3pt}                                                                                                                                                                                                                                                                                                                                                                                                                                                                                                                                                     \\

        P06                                                                                                           & \tikzcircleblack{3pt} \tikzcirclewhite{3pt} \tikzcirclewhite{3pt} \tikzcirclewhite{3pt} \tikzcirclewhite{3pt} &
        \tikzcircleblack{3pt} \tikzcircleblack{3pt} \tikzcircleblack{3pt} \tikzcircleblack{3pt} \tikzcirclewhite{3pt} &
        \tikzcircleblack{3pt} \tikzcircleblack{3pt} \tikzcircleblack{3pt} \tikzcirclewhite{3pt} \tikzcirclewhite{3pt} &
        \tikzcircleblack{3pt} \tikzcircleblack{3pt} \tikzcircleblack{3pt} \tikzcirclewhite{3pt} \tikzcirclewhite{3pt} &
        \tikzcircleblack{3pt} \tikzcircleblack{3pt} \tikzcircleblack{3pt} \tikzcirclewhite{3pt} \tikzcirclewhite{3pt} &
        \tikzcircleblack{3pt} \tikzcircleblack{3pt} \tikzcircleblack{3pt} \tikzcirclewhite{3pt} \tikzcirclewhite{3pt} &
        \tikzcircleblack{3pt} \tikzcircleblack{3pt} \tikzcircleblack{3pt} \tikzcirclewhite{3pt} \tikzcirclewhite{3pt}                                                                                                                                                                                                                                                                                                                                                                                                                                                                                                                                                     \\

        P07                                                                                                           & \tikzcircleblack{3pt} \tikzcirclewhite{3pt} \tikzcirclewhite{3pt} \tikzcirclewhite{3pt} \tikzcirclewhite{3pt} &
        \tikzcircleblack{3pt} \tikzcircleblack{3pt} \tikzcircleblack{3pt} \tikzcirclewhite{3pt} \tikzcirclewhite{3pt} &
        \tikzcircleblack{3pt} \tikzcircleblack{3pt} \tikzcirclewhite{3pt} \tikzcirclewhite{3pt} \tikzcirclewhite{3pt} &
        \tikzcircleblack{3pt} \tikzcirclewhite{3pt} \tikzcirclewhite{3pt} \tikzcirclewhite{3pt} \tikzcirclewhite{3pt} &
        \tikzcircleblack{3pt} \tikzcircleblack{3pt} \tikzcirclewhite{3pt} \tikzcirclewhite{3pt} \tikzcirclewhite{3pt} &
        \tikzcircleblack{3pt} \tikzcircleblack{3pt} \tikzcirclewhite{3pt} \tikzcirclewhite{3pt} \tikzcirclewhite{3pt} &
        \tikzcircleblack{3pt} \tikzcirclewhite{3pt} \tikzcirclewhite{3pt} \tikzcirclewhite{3pt} \tikzcirclewhite{3pt}                                                                                                                                                                                                                                                                                                                                                                                                                                                                                                                                                     \\

        P08                                                                                                           & \tikzcircleblack{3pt} \tikzcirclewhite{3pt} \tikzcirclewhite{3pt} \tikzcirclewhite{3pt} \tikzcirclewhite{3pt} &
        \tikzcircleblack{3pt} \tikzcircleblack{3pt} \tikzcircleblack{3pt} \tikzcirclewhite{3pt} \tikzcirclewhite{3pt} &
        \tikzcircleblack{3pt} \tikzcirclewhite{3pt} \tikzcirclewhite{3pt} \tikzcirclewhite{3pt} \tikzcirclewhite{3pt} &
        \tikzcircleblack{3pt} \tikzcircleblack{3pt} \tikzcirclewhite{3pt} \tikzcirclewhite{3pt} \tikzcirclewhite{3pt} &
        \tikzcircleblack{3pt} \tikzcircleblack{3pt} \tikzcirclewhite{3pt} \tikzcirclewhite{3pt} \tikzcirclewhite{3pt} &
        \tikzcircleblack{3pt} \tikzcirclewhite{3pt} \tikzcirclewhite{3pt} \tikzcirclewhite{3pt} \tikzcirclewhite{3pt} &
        \tikzcircleblack{3pt} \tikzcirclewhite{3pt} \tikzcirclewhite{3pt} \tikzcirclewhite{3pt} \tikzcirclewhite{3pt}                                                                                                                                                                                                                                                                                                                                                                                                                                                                                                                                                     \\

        P09                                                                                                           & \tikzcircleblack{3pt} \tikzcirclewhite{3pt} \tikzcirclewhite{3pt} \tikzcirclewhite{3pt} \tikzcirclewhite{3pt} &
        \tikzcircleblack{3pt} \tikzcircleblack{3pt} \tikzcirclewhite{3pt} \tikzcirclewhite{3pt} \tikzcirclewhite{3pt} &
        \tikzcircleblack{3pt} \tikzcircleblack{3pt} \tikzcircleblack{3pt} \tikzcirclewhite{3pt} \tikzcirclewhite{3pt} &
        \tikzcircleblack{3pt} \tikzcircleblack{3pt} \tikzcirclewhite{3pt} \tikzcirclewhite{3pt} \tikzcirclewhite{3pt} &
        \tikzcircleblack{3pt} \tikzcircleblack{3pt} \tikzcirclewhite{3pt} \tikzcirclewhite{3pt} \tikzcirclewhite{3pt} &
        \tikzcircleblack{3pt} \tikzcircleblack{3pt} \tikzcircleblack{3pt} \tikzcirclewhite{3pt} \tikzcirclewhite{3pt} &
        \tikzcircleblack{3pt} \tikzcirclewhite{3pt} \tikzcirclewhite{3pt} \tikzcirclewhite{3pt} \tikzcirclewhite{3pt}                                                                                                                                                                                                                                                                                                                                                                                                                                                                                                                                                     \\

        P10                                                                                                           & \tikzcircleblack{3pt} \tikzcircleblack{3pt} \tikzcircleblack{3pt} \tikzcirclewhite{3pt} \tikzcirclewhite{3pt} &
        \tikzcircleblack{3pt} \tikzcircleblack{3pt} \tikzcircleblack{3pt} \tikzcirclewhite{3pt} \tikzcirclewhite{3pt} &
        \tikzcircleblack{3pt} \tikzcircleblack{3pt} \tikzcirclewhite{3pt} \tikzcirclewhite{3pt} \tikzcirclewhite{3pt} &
        \tikzcircleblack{3pt} \tikzcirclewhite{3pt} \tikzcirclewhite{3pt} \tikzcirclewhite{3pt} \tikzcirclewhite{3pt} &
        \tikzcircleblack{3pt} \tikzcircleblack{3pt} \tikzcircleblack{3pt} \tikzcirclewhite{3pt} \tikzcirclewhite{3pt} &
        \tikzcircleblack{3pt} \tikzcircleblack{3pt} \tikzcircleblack{3pt} \tikzcirclewhite{3pt} \tikzcirclewhite{3pt} &
        \tikzcircleblack{3pt} \tikzcircleblack{3pt} \tikzcirclewhite{3pt} \tikzcirclewhite{3pt} \tikzcirclewhite{3pt}                                                                                                                                                                                                                                                                                                                                                                                                                                                                                                                                                     \\

        P11                                                                                                           & \tikzcircleblack{3pt} \tikzcirclewhite{3pt} \tikzcirclewhite{3pt} \tikzcirclewhite{3pt} \tikzcirclewhite{3pt} &
        \tikzcircleblack{3pt} \tikzcircleblack{3pt} \tikzcircleblack{3pt} \tikzcirclewhite{3pt} \tikzcirclewhite{3pt} &
        \tikzcircleblack{3pt} \tikzcircleblack{3pt} \tikzcirclewhite{3pt} \tikzcirclewhite{3pt} \tikzcirclewhite{3pt} &
        \tikzcircleblack{3pt} \tikzcircleblack{3pt} \tikzcirclewhite{3pt} \tikzcirclewhite{3pt} \tikzcirclewhite{3pt} &
        \tikzcircleblack{3pt} \tikzcircleblack{3pt} \tikzcircleblack{3pt} \tikzcircleblack{3pt} \tikzcirclewhite{3pt} &
        \tikzcircleblack{3pt} \tikzcircleblack{3pt} \tikzcircleblack{3pt} \tikzcirclewhite{3pt} \tikzcirclewhite{3pt} &
        \tikzcircleblack{3pt} \tikzcircleblack{3pt} \tikzcircleblack{3pt} \tikzcirclewhite{3pt} \tikzcirclewhite{3pt}                                                                                                                                                                                                                                                                                                                                                                                                                                                                                                                                                     \\

        P12                                                                                                           & \tikzcircleblack{3pt} \tikzcirclewhite{3pt} \tikzcirclewhite{3pt} \tikzcirclewhite{3pt} \tikzcirclewhite{3pt} &
        \tikzcircleblack{3pt} \tikzcircleblack{3pt} \tikzcircleblack{3pt} \tikzcirclewhite{3pt} \tikzcirclewhite{3pt} &
        \tikzcircleblack{3pt} \tikzcirclewhite{3pt} \tikzcirclewhite{3pt} \tikzcirclewhite{3pt} \tikzcirclewhite{3pt} &
        \tikzcircleblack{3pt} \tikzcirclewhite{3pt} \tikzcirclewhite{3pt} \tikzcirclewhite{3pt} \tikzcirclewhite{3pt} &
        \tikzcircleblack{3pt} \tikzcircleblack{3pt} \tikzcircleblack{3pt} \tikzcirclewhite{3pt} \tikzcirclewhite{3pt} &
        \tikzcircleblack{3pt} \tikzcircleblack{3pt} \tikzcircleblack{3pt} \tikzcirclewhite{3pt} \tikzcirclewhite{3pt} &
        \tikzcircleblack{3pt} \tikzcircleblack{3pt} \tikzcircleblack{3pt} \tikzcirclewhite{3pt} \tikzcirclewhite{3pt}                                                                                                                                                                                                                                                                                                                                                                                                                                                                                                                                                     \\ \bottomrule
    \end{tabular}
\end{table*}

Next, participants received a 5-minute overview of SliceViz.
Since we wanted them to try out the tool themselves, we skipped a detailed demo.

Participants were then asked to access SliceViz through a web server and complete three tasks:

\textbf{Task 1:} Participants analyzed the \emph{Button3} app from the DroidBench micro-benchmark~\cite{droidbench}. Their goal was to identify the data collected, determine GDPR protection requirements, and observe privacy-relevant properties in the resulting program slice. While the dashboard provided partial support, fully assessing privacy-relevant properties required further inspection of the data flow in the visualized program slices.

% \textbf{Task 2:} Participants performed the same analysis as in Task 1, but on the more complex godwon\_samp.apk app from the TaintBench suite~\cite{taintbench}.

% \textbf{Task 3:} Participants were asked to compare the usability, simplicity, and relevance of the Jimple and Java Views for the Button3.apk app (Task 1 app).

\textbf{Task 2:} Participants were asked to compare the usability, simplicity, and relevance of the Jimple and Java Views for the \emph{Button3} app (the same app given in Task 1).

\textbf{Task 3:} Participants performed the same analysis as in Task 1, but on the more complex \emph{godwon\_samp} app from the TaintBench suite~\cite{taintbench}.

Tasks 1 and 3 were designed to assess if SliceViz assists developers in understanding data protection in Android apps (\textit{RQ1}), and were conducted on two distinct apps to identify data flows and privacy properties. 
Ground truth information was available for both applications, and participant responses were manually reviewed against the documented ground truth and the source code to ensure consistency and accuracy.

Task 2 was designed to evaluate how effectively the experimental Java View aids in understanding program slices compared to the Jimple View (\textit{RQ2}). To assess this, participants explored both views using a DroidBench app, enabling a direct comparison of their effectiveness in supporting program understanding.

After completing all tasks, participants answered several 5-point Likert-type questions from the System Usability Scale (SUS)~\cite{sus}, a questionnaire designed to measure the effectiveness and efficiency of a system, and rated SliceViz using a Net Promoter Score (NPS)~\cite{nps}, that measures their likelihood of recommending the tool to a friend.

\subsection{Results}

%\mk{We will have at least two more participants, who have already registered for the user study in Feb}

In this section, we present the results of the user study conducted with 12 participants, who we refer to as P01–P12.
We discarded P05's data, since they declared themselves to be a beginner and novice in all skills, and required additional assistance and time during the study, far beyond what was needed by other participants.
We summarize the results below:

\textbf{Participant Experience.} At the beginning of the study, participants self-reported familiarity with Android app development (median 1/5), Java (median 3/5), Jimple (median 2/5), static analysis tools (median 2/5), graph visualization (median 3/5), data privacy (median 3/5), and GDPR (median 3/5) (cf.~\Cref{tab:participants}).
Participants have a median of approximately 1.5 years of experience in Java development, with maximum experience of 5 years.
%experiments on a Linux machine with Ubuntu 20.04.2, an Intel i7-10850H 6-core processor and 32 GB memory.
%We first ran PriBaSE and SliceViz and tested their functionality. 
%Next, we performed two pilot studies to investigate the usability of SliceViz and its usefulness for developers, respectively.
%Based on our findings, we further contribute another optimization.

\textbf{RQ1. To what extent does SliceViz help developers understand data protection in Android apps?}

Participants first worked with a DroidBench app, which had one privacy-relevant program slice with 10 nodes.
Despite limited knowledge of GDPR, 10 participants successfully identified the privacy risks in the app and accurately determined which data required GDPR protection.
Apart from P04 who over-reported the risky sources, all the other participants were able to analyze the slices, interpret the graphs, and understand the data flow, correctly identifying privacy-relevant properties.
\begin{comment}
\eb{I do not understand what exactly the task was. I mean SliceViz already does a lot of what we ask the users to do. It identifies and classifies sources, etc. So what is the additional work that participants needed to perform? And what were mistakes that they could actually make there?}
\mk{Discuss in JF.}
\end{comment}

The TaintBench app given for Task~3 had seven program slices, with a maximum size of 20 nodes.
Of the 11 participants, 10 accurately identified all the privacy-relevant information in the given app.
P03 correctly identified 50\% of the privacy-relevant details, but failed to report risk 4 slices, likely due to customizing SliceViz to only report risk 1 data sources.
Interestingly, P04 (who struggled to understand privacy risks and data flow in the first task) could successfully identify and report the privacy-relevant details in this app, suggesting a slight learning curve.

\begin{figure*}[t]
    \begin{center}
        \includegraphics[width=0.8\textwidth]{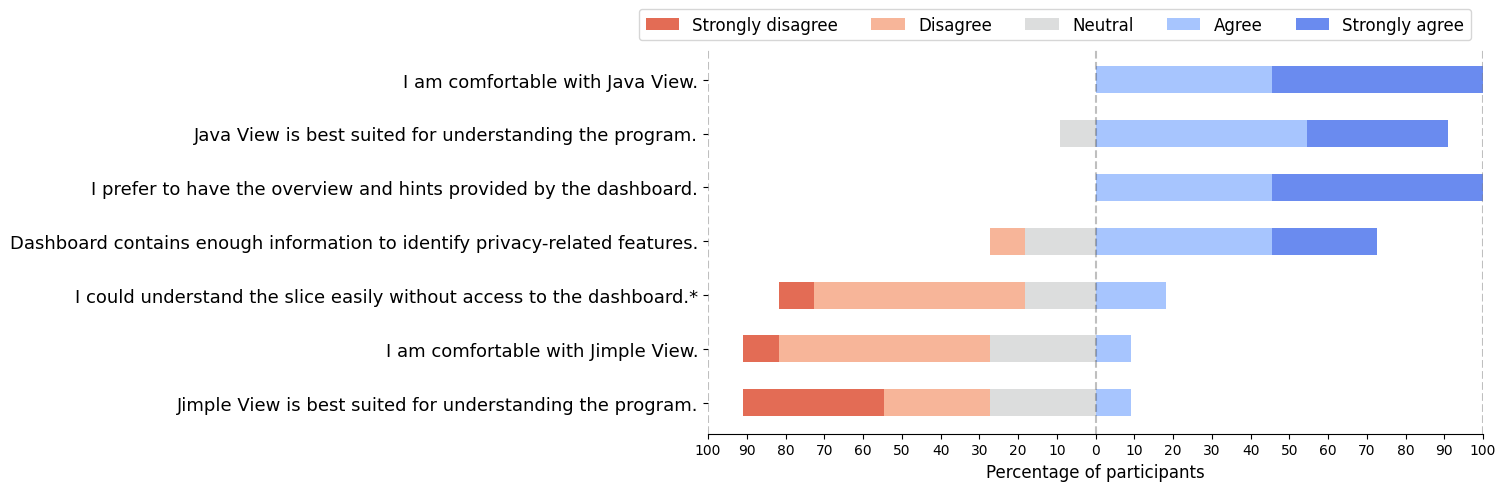}
        \caption{Participants' responses to the tool questions. For the statement marked with $\ast$, \emph{Strongly Disagree} represents a more favorable response.}
        \label{fig:divergedstackbar}
    \end{center}
\end{figure*}

% The first two tasks included questions about the dashboard.
The questionnaire also included dashboard evaluation questions, which helped assess how effectively the dashboard supported participants in understanding data protection aspects within Android apps.
All 11 participants agreed that the overview and the hints provided by the dashboard helped them better understand the tool (cf.~\Cref{fig:divergedstackbar}).
P02 suggested more detailed description of all data sources within the dashboard.

Four participants \textit{(P03, P07, P10, and P12)} explicitly mentioned the warning scale as one of the most useful features of the dashboard.
P03 noted, \textit{``I like how all risks are categorized with source and warning scale''}.
P10 said, \textit{``Warning scale is the best way to convey overall risks''}.
P04 found the bar and pie charts easy to understand.
P09 said, \textit{``SliceViz gave me a good understanding about privacy-relevant data flows. Warning scale was helpful. I prioritized looking at slices rated D before I moved on to safer slices''}.
P12 confessed that SliceViz made them aware of potential privacy risks.

\begin{comment}
\stilltodo{TODO: Wondering if we should represent Likert scale data using a percentage stacked bar chart. Such as Figure~\ref{divergedstackbar}.}
\mk{I personally don't think it gives too much information, and we may not have enough space for it in the paper. But I can make space if @EB and @MS like the idea of displaying this chart.}
\ms{Having the data is nice and I do understand Figure~\ref{divergedstackbar}, but I do not need it. If it is in the artifact that would be good enough for me - also tex	t reference.}
\mk{Will add Figure~\ref{divergedstackbar} to text if we decide to keep it in the paper. What about Figure~\ref{fig:npsviews}? Should we keep that figure too, or can I remove it?}
\eb{I think Figure 8 does not hurt but it is also not needed. If we need the space then you can drop it.}
\end{comment}

\begin{comment}
\eb{The following scores, are they for the Jimple or Java version? If it's for the overall tool, irrespective of the view, we should clarify this.}
\mk{So this RQ has nothing to do with the different views. I now clarified this in RQ2.}
\end{comment}

Using the System Usability Scale (SUS), SliceViz obtained a mean score of 63.96, which falls below the industry benchmark of 68.
Notably, users with no prior Java development experience expressed dissatisfaction with its usability.
%This lower score may be attributed to the fact that SliceViz is the result of an initial iteration in the user-centered design process. 
Since SliceViz is designed for privacy analysis and visualization, the gap in experience with Android development and static analysis tools (cf.~\Cref{tab:participants}) could explain some of the usability challenges reported by less experienced users.

However, participants with one to five years of Java development experience rated SliceViz with an average SUS score of 68, indicating a satisfactory level of usability.  These findings suggest that SliceViz provides an acceptable usability experience for Java and Android developers.
% who do not have an with intermediate experience in privacy analysis or graph visualization tools.

The Net Promoter Score (NPS) for SliceViz was calculated as 18.18, reflecting a positive perception among participants.\\

%\textbf{Other feedback.} 
\begin{mdframed}[backgroundcolor=black!10,roundcorner=8pt]
    \textbf{\textbf{RQ1.}} SliceViz effectively assists developers with prior development experience in identifying and observing privacy-relevant properties in Android apps.
\end{mdframed}

\begin{comment}
\begin{figure}[t]
    \begin{center}
        \includegraphics[width=0.5\textwidth]{Figures/RatingViews1.pdf}
        \caption{Rating of Jimple and Java Views. (0: not recommended, 10: highly recommended)}%\ms{Do we know why P09 preferred Jimple?}\mk{Yes, they thought Jimple view was more detailed and hence gave it a good rating, but also acknowledged that it is complicated for an average Java developer.}
        \label{fig:npsviews}
    \end{center}
\end{figure}
\end{comment}

\textbf{RQ2. To what extent does the experimental Java View better support developers in understanding the flow of privacy-relevant data within an app?}

Most of the participants (N=10) agreed that the Java View was \textbf{easier} and \textbf{faster to understand} compared to the Jimple View: \textit{``I have more experience with Java so I could quickly understand the Java View''} (P07), \textit{``Unlike Jimple View, Java view shows the data of risk sources in a simple way instead of giving too many details.''} (P02).
While P09 also found the Java View quicker to grasp, they appreciated the Jimple View for its detailed control flow information.
However, they acknowledged that their familiarity with Jimple made it easier to follow, and those less experienced with Jimple might struggle with its unfamiliar syntax and temporary variables.
Notably, all 11 participants agreed or strongly agreed that they were \textbf{more comfortable} with Java View (cf.~\Cref{fig:divergedstackbar}).

To compare user preferences for both views, participants were asked, \textit{``How likely is it you would recommend Java View over Jimple View to a friend?''} (and vice versa).
Their responses were used to calculate separate Net Promoter Scores (NPS) for each view.
The NPS for Java View was 45.45, reflecting strong user preference over the Jimple View.
In contrast, the Jimple View received a negative score of -72.73. % (individual ratings in \Cref{fig:npsviews}).  

Participants found the Jimple View \textbf{complex} (P06: \textit{``The Jimple View has too many
    nodes and edges''}), \textbf{hard to understand,} and \textbf{not user-friendly for beginners} (P02: \textit{``It is not understandable for any non-programmer or normal user''}).
P10 noted, \textit{``I don't know the Jimple syntax.''}.
P11 complained, \textit{``The Jimple View contains so many other constructs and naming techniques that require explicit familiarity''}.

On the other hand, the Java View received good feedback.
For instance, P03 commented, \textit{``The Java View aligns with the source code I’m familiar with in Android Studio,''} showcasing how the representation resonates with developers accustomed to Java environments.
P07 stated, \textit{``The Java View captures the necessary flow without overwhelming the user with too much detail.''}
P01 claimed the Java View was \textit{coherent with source code}.
P11 said, \textit{``It's much more concise and helps me focus more on the actual code that I'm familiar with.''}
P12 thought that Java View was \textit{human readable, and less cluttered.}
%\textbf{Other feedback.} 

\textbf{Overall feedback.} The final set of questions had some open questions asking for feedback.

Participants generally regarded SliceViz as a \textit{good tool for developers} (P01) which was \textit{reasonably easy to use} (P01), with some describing it as a \textit{promising start} (P03) and noting its \textit{great usability} (P02).
Ten out of eleven participants found the documentation helpful, and P02 appreciated the \textit{inclusion of relevant research papers linked within the documentation}.
P03 highlighted the benefit of being able to access all previous results.
P08 appreciated being able to track the whole workflow with helpful legends.

Participants also offered valuable suggestions for improving SliceViz.
P04 recommended providing more information about Jimple and suggested in-depth modifications to the user interface.
P06 pointed out that graphical views may not be very effective for handling large codebases.
P03 proposed including more specific hints to clarify the meaning of the graph views.\\
%P09 and P10 started the study assuming that SliceViz identifies security vulnerabilities. 
%After a short discussion, they appreciated how SliceViz does not \textit{just} identify leaks, but also reasons about data protection. 

\begin{mdframed}[backgroundcolor=black!10,roundcorner=8pt]
    \textbf{\textbf{RQ2.}} Developers are comfortable with Java View, since it is more concise and aligns with the source code.
\end{mdframed}

%% file: Sections/3_RelatedWork.tex
\section{Related Work}
\label{relatedwork}

%\eb{Must make sure to explain why these works do not sufficiently address the research gap.}
%\mk{Wondering if it is better to have related work \emph{after} explaining the tool. So we can refer to some tables to explicitly say how our tool weaves in GDPR articles in the analysis.}
%\ms{Add A-Mobile Papers - check with Eric for Assesor View <-> conflict?}
%\mk{Added the PRICE paper. Don't understand where assessor view would fit.}
%\ms{I thought of Assesor View as it builds on this paper as related work. However, citing it could unblind this paper. Thus, the mentioned conflict. Not referencing it thus makes sense.}
%\mk{SCAM reviewers suggested pushing related work before introducing the approach. Does this read well?}
%We next discuss existing research on this topic. 
We discuss prior research in three major areas relevant to our work: (1) program slicing and visualization, (2) data protection assessment, and (3) data safety section.

%Goal 3: Unified framework to assist technical and legal experts
\textbf{Program slicing and visualization.} Since we focus on analyzing Android apps, we concentrate on slicing techniques tailored to the Android ecosystem. 
While program slicing has a rich history across platforms~\cite{surveyslicing}, Jicer~\cite{jicer} is the only available static slicer for Android apps.
It uses Soot~\cite{soot}'s DotGraph API~\cite{dotgraph} to generate static graphs in SVG format. 
However, these graphs do not scale well and lack interactivity. 
Several dynamic slicing approaches for Android have also been proposed~\cite{dynamicslicing,mandoline}. 

In 2007, Sridharan et al.~\cite{thinslicing} proposed WALA~\cite{wala}-based thin slicing, which excludes control dependencies and focuses only on statements that compute and copy the value returned by the slicing criterion—a strategy relevant to our compliance goals. 
Prior work has visualized program slices either graphically~\cite{dynamicviz, visualizeslices} or textually~\cite{textslices}, but these visualizations were primarily designed for debugging and program understanding. %, not for privacy compliance.
Our work builds on these foundations but shifts the focus from traditional debugging or general program understanding toward compliance-oriented program comprehension and analysis. 
%Specifically, we use static program slicing that preserves both control and data dependencies, and we develop a visualization interface tailored to identifying privacy risks in accordance with GDPR obligations. 

\textbf{Data protection assessment.} A growing body of work investigates how static analysis techniques can support privacy risk assessment. 
Tang and Østvold~\cite{privflow} describe an automatic source-to-sink analysis technique that presents the results as a graph of privacy flows and operations to support app developers and DPOs in evaluating privacy compliance and simplifying DPIAs. 
However, their approach focuses on a limited set of data sources—namely, database, network, and I/O—which may miss many relevant types of personal data, particularly those covered under the GDPR. 
%\eb{Are there any concrete deficiencies of the related work that we address? Would be good to have some.}
%\mk{Addressed.}

Similarly, other taint analysis-based tools~\cite{PTPDroid,atpchecker,privacycat} use FlowDroid~\cite{flowdroid} to analyze data flows, primarily to detect privacy vulnerabilities, or verify third-party library compliance.  
While effective at identifying direct leaks, traditional taint analysis is limited in scope~\cite{khedkarmobilesoft,ferrara}. 
It operates by analyzing reachability between predefined sources (where sensitive data originates) and sinks (where data may be exposed). 
Without known \emph{predefined} sinks, taint analysis cannot proceed.

Our approach overcomes this limitation. 
Rather than requiring predefined sinks, we take known sources as input and identify potential sinks as output. 
While, in theory, taint analysis could be adapted for GDPR compliance by treating every program point as a potential sink, this is not scalable in practice.  
%We conduct a source-centric, program slicing-based analysis that does not depend on a predefined set of sinks. 
In contrast, our source-centric, slicing-based analysis supports broader coverage and enables a more nuanced, risk-based assessment of personal data handling.
It supports a more actionable evaluation of GDPR obligations (cf.~\Cref{tab:WarningScale}).  %such as data minimization~\cite{art4}

\textbf{Data safety section.} Since its introduction in 2022, the Data Safety Section (DSS) has received increasing attention from researchers due to its mandatory status on the Google Play Store~\cite{data}. 
A Mozilla study~\cite{mozilla} graded popular apps based on discrepancies between data safety sections and privacy policies. 
Khandelwal et al.~\cite{datalabels} noted underreporting, overreporting, and inconsistencies in data practices, analyzing developer interactions using snapshot data without source code analysis. 
In contrast, our approach uses static code analysis, enabling more precise and actionable insights to support developers in accurately reporting their data practices. 

Google’s Checks~\cite{googlechecks} is a paid service that helps developers complete the Data Safety Section, but its cost may be prohibitive. 
Matcha~\cite{matcha}, an open-source JetBrains plugin, analyzes source code to suggest labels, requiring developers to annotate their code. 
In contrast, our approach works directly on APKs without needing annotations and goes beyond label generation by identifying privacy-relevant data and providing timely GDPR-specific warnings (cf.~\Cref{tab:WarningScale})—an aspect current tools do not address. 
%\eb{It seems that many basic services are free. Do we have other arguments? https://checks.google.com/pricing/}
%\mk{Addressed.}

Our recent experiment~\cite{asej_price} has revealed discrepancies between the data collection reported in data safety sections and the source code of the app, highlighting the need for improved tool support to aid Android developers in understanding legal privacy implications in their source code. 
%Their approach introduces a prototype to statically detect and label the privacy-relevant data collected by the source code, user interface screens, and permissions. 
%Their findings 
SliceViz builds on this need by combining program slicing, visualization, and compliance-aware analysis to help developers understand and manage legal privacy implications at a deeper level. %---a gap that we aim to address with SliceViz.

In summary, while substantial research exists on data privacy and visualization of program slicing, our approach is the first to integrate program slicing, visualization, and risk-based privacy assessment to directly support GDPR compliance. 

%% file: Sections/7_Limitations.tex
\section{Limitations and Threats to Validity}
\label{limitations}
We next discuss the limitations of our approach.

\textbf{Analysis.} The analysis at the back-end of SliceViz uses the \textit{Identifier Dataset}, which may miss out on some libraries. 
However, we ensure a comprehensive evaluation by including the most used third-party libraries from different categories~\cite{asej_price}. 
The degree of identifiability relies on human subjectivity and it may vary with different domains and compliance regimes. 
Since our approach focuses on GDPR, we can use the \textit{Identifier Dataset} directly. 
The Slicing Engine matches the method signatures in the ADG with the labeled methods present in the \textit{Identifier Dataset}. 
Since we use keyword matching to detect the presence of data sources including those from some third-party libraries, the analysis cannot handle code obfuscation in cases where the obfuscation also renames calls to these third-party APIs. 
However, as mentioned in \Cref{impl}, the Slicing Engine is designed to operate as a proactive privacy measure before name mangling or obfuscation measures are applied. 
The Java View can be improved by expanding the mapping between Jimple and Java, comprehensively covering the source code. 
%Scalability

\textbf{Usability Assessment.} To investigate possible usability implications of SliceViz, we conducted two pilot studies and one user study. 
Since it is challenging to recruit developers for a user study and students are considered reasonable proxies for software developers~\cite{userstudy1,userstudy2,userstudy3}, our user study used students as participants. 
%While the participant quantity was relatively small, the qualitative feedback and the results from the NPS and SUS still provide valuable insights. 
%\eb{What does this mean do we or don't we have statistically significant results? Did you test for that?}
We conducted the user study in a controlled environment (12 participants, 1 hour study duration) rather than in a development setting. 
In practice, users would have more time to investigate more advanced apps. 
It would, however, be interesting to conduct a larger quantitative evaluation in real-life conditions.
%\ms{Address low number of participants for NPS and SUS, however they still maybe an indicator.}

%% file: Sections/8_Conclusion.tex
\section{Conclusion}
\label{conclusion}

In this paper, we have explored how static program slicing can be used to identify parts of the source code affected by sources of privacy-relevant data.

To achieve this, we have introduced \emph{SliceViz}, a web tool that analyzes an Android app by slicing all privacy-relevant data sources detected in the source code. 
It then helps developers by visualizing these privacy-relevant program slices. 

\begin{comment}
We have evaluated PriBaSE and SliceViz on 36 applications: 5 apps from DroidBench, 25 apps from TaintBench, and 6 real-world Android apps available on Google Play Store. 
The smaller slices visualized by SliceViz provide valuable insights into privacy-relevant data collection and logging, but larger graphs caused information overload. 
Our studies suggest that the visualization of SliceViz helps developers in understanding flows of privacy-relevant data in applications, however, users recommend more contextual information. 
Moreover, users unfamiliar with Jimple struggle to interpret SliceViz's graphs. 
We have observed that the slices generated for large real-world apps, although not ideal for visualization, contain evidence of data protection measures. Further, we have observed that excluding control dependencies leads to a significant decrease in the sizes of privacy-relevant program slices. 
This optimized version of PriBaSE helps SliceViz in reducing visual load. 

Finally, we have presented usability improvements that address two key issues users experienced with our approach: the lack of contextual information, and the complexity of the Jimple view. 
\end{comment}

With a user study conducted on 12 participants, we have shown that SliceViz effectively assists developers with prior development experience in identifying and observing privacy-relevant properties in Android apps. 

Our findings suggest that program slicing can be employed to reason about privacy-relevant data flows in Android applications. 
With further usability improvements, developers can be better equipped to handle privacy-sensitive information.
%\ms{In approach and implementation I was not always sure what our contribution in this paper is and what came from Khedkar et al. MAybe we can make this cleare?}
%\eb{I do not see this problem. The earlier paper had no slicing and no visualization, right? And the mobilesoft paper just mentioned the idea but not more.}

%% file: sample-base.bib
@article{assessorview,
  author  = {Khedkar, Mugdha and Schlichtig, Michael and Atakishiyev, Nihad and Bodden, Eric},
  title   = {Between Law and Code: Challenges and Opportunities for Automating Privacy Assessments},
  journal = {Automated Software Engineering},
  year    = {2026},
  volume  = {33},
  number  = {56},
  pages   = {45},
  doi     = {10.1007/s10515-026-00601-4},
  url     = {https://doi.org/10.1007/s10515-026-00601-4},
  issn    = {1573-7535}
}

@misc{appbrain,
  year =         "2014",
  title =        "AppBrain",
  url =         "https://www.appbrain.com/",
  lastaccessed = "May 28, 2024",
}

@misc{withinsubjects,
  year =         "2023",
  title =        "Nielsen Normal Group. Between-subjects vs. within-subjects study design.",
  url =         "https://www.nngroup.com/articles/between-within-subjects/",
  lastaccessed = "March 6, 2025",
}

@misc{gdpr,
  year =         "2018",
  title =        "The European parliament and the council of the European union. General Data Protection Regulation (GDPR)",
  url =          "https://eur-lex.europa.eu/legal-content/EN/TXT/PDF/?uri=CELEX:32016R0679",
  lastaccessed = "May 13, 2024",
}

@misc{mozilla,
  year =         "2023",
  title =        "See No Evil: Loopholes in Google’s Data Safety Labels Keep Companies in the Clear and Consumers in the Dark",
  url =          "https://foundation.mozilla.org/en/campaigns/googles-data-safety-labels/",
  lastaccessed = "May 28, 2024",
}

@misc{data,
  year =         "2022",
  title =        "Data Safety Section",
  url =         "https://blog.google/products/google-play/data-safety/",
  lastaccessed = "May 28, 2024",
}

@misc{d3.js,
   year = "2024",
   title = "D3.js Library",
   url = "https://d3js.org/",
   lastaccessed = "June 15, 2024",
}

@misc{su2025codestatementsimplementprivacy,
      title={Which Code Statements Implement Privacy Behaviors in Android Applications?}, 
      author={Chia-Yi Su and Aakash Bansal and Vijayanta Jain and Sepideh Ghanavati and Sai Teja Peddinti and Collin McMillan},
      year={2025},
      eprint={2503.02091},
      archivePrefix={arXiv},
      primaryClass={cs.SE},
      url={https://arxiv.org/abs/2503.02091}, 
}

@ARTICLE{d3paper,
  author={Bostock, Michael and Ogievetsky, Vadim and Heer, Jeffrey},
  journal={IEEE Transactions on Visualization and Computer Graphics}, 
  title={D³ Data-Driven Documents}, 
  year={2011},
  volume={17},
  number={12},
  pages={2301-2309},
  doi={10.1109/TVCG.2011.185}}

@misc{niroshan2025empiricalstudycodeobfuscation,
      title={An Empirical Study of Code Obfuscation Practices in the Google Play Store}, 
      author={Akila Niroshan and Suranga Seneviratne and Aruna Seneviratne},
      year={2025},
      eprint={2502.04636},
      archivePrefix={arXiv},
      primaryClass={cs.CR},
      url={https://arxiv.org/abs/2502.04636}, 
}

@inproceedings{obfuscationuse,
author = {Wermke, Dominik and Huaman, Nicolas and Acar, Yasemin and Reaves, Bradley and Traynor, Patrick and Fahl, Sascha},
title = {A Large Scale Investigation of Obfuscation Use in Google Play},
year = {2018},
isbn = {9781450365697},
publisher = {Association for Computing Machinery},
address = {New York, NY, USA},
url = {https://doi.org/10.1145/3274694.3274726},
doi = {10.1145/3274694.3274726},
booktitle = {Proceedings of the 34th Annual Computer Security Applications Conference},
pages = {222–235},
numpages = {14},
location = {San Juan, PR, USA},
series = {ACSAC '18}
}

@inproceedings{privacycat,
author = {Mao, Ke and \r{A}hs, Cons and Cela, Sopot and Distefano, Dino and Gardner, Nick and Grigore, Radu and Gustafsson, Per and Hajdu, \'{A}kos and Kapus, Timotej and Marescotti, Matteo and Sampaio, Gabriela Cunha and Suzanne, Thibault},
title = {PrivacyCAT: Privacy-Aware Code Analysis at Scale},
year = {2024},
isbn = {9798400705014},
publisher = {Association for Computing Machinery},
address = {New York, NY, USA},
url = {https://doi.org/10.1145/3639477.3639742},
doi = {10.1145/3639477.3639742},
booktitle = {Proceedings of the 46th International Conference on Software Engineering: Software Engineering in Practice},
pages = {106–117},
numpages = {12},
location = {Lisbon, Portugal},
series = {ICSE-SEIP '24}
}

@inproceedings{ferrara,
author = {Ferrara, Pietro and Spoto, Fausto},
year = {2018},
month = {01},
pages = {},
title = {Static Analysis for GDPR Compliance}
}

@article{sus,
author = {Rebecca A. Grier and Aaron Bangor and Philip Kortum and S. Camille Peres},
title ={The System Usability Scale: Beyond Standard Usability Testing},

journal = {Proceedings of the Human Factors and Ergonomics Society Annual Meeting},
volume = {57},
number = {1},
pages = {187-191},
year = {2013},
doi = {10.1177/1541931213571042},

URL = { 
    
        https://doi.org/10.1177/1541931213571042
    
  

},
eprint = { 
    
        https://doi.org/10.1177/1541931213571042
    
    

}
}

@inproceedings{avproposal,
author = {Khedkar, Mugdha and Schlichtig, Michael and Bodden, Eric},
title = {Advancing Android Privacy Assessments with Automation},
year = {2024},
isbn = {9798400712494},
publisher = {Association for Computing Machinery},
address = {New York, NY, USA},
url = {https://doi.org/10.1145/3691621.3694953},
doi = {10.1145/3691621.3694953},
booktitle = {Proceedings of the 39th IEEE/ACM International Conference on Automated Software Engineering Workshops},
pages = {218–222},
numpages = {5},
location = {Sacramento, CA, USA},
series = {ASEW '24}
}

@INPROCEEDINGS{visualizeslices,
  author={Ball, T. and Eick, S.G.},
  booktitle={Proceedings of 1994 IEEE Symposium on Visual Languages}, 
  title={Visualizing program slices}, 
  year={1994},
  volume={},
  number={},
  pages={288-295},
  doi={10.1109/VL.1994.363606}}

@INPROCEEDINGS{textslices,
  author={Krinke, J.},
  booktitle={20th IEEE International Conference on Software Maintenance, 2004. Proceedings.}, 
  title={Visualization of program dependence and slices}, 
  year={2004},
  volume={},
  number={},
  pages={168-177},
  doi={10.1109/ICSM.2004.1357801}}

@INPROCEEDINGS {buggingyousomuch,
author = { Horstmann, Stefan Albert and Hong, Sandy and Klein, David and Serafini, Raphael and Degeling, Martin and Johns, Martin and Moonsamy, Veelasha and Naiakshina, Alena },
booktitle = { 2025 IEEE Symposium on Security and Privacy (SP) },
title = {{ “Sorry for bugging you so much.” Exploring Developers’ Behavior Towards Privacy-Compliant Implementation }},
year = {2025},
volume = {},
ISSN = {2375-1207},
pages = {1215-1233},
doi = {10.1109/SP61157.2025.00146},
url = {https://doi.ieeecomputersociety.org/10.1109/SP61157.2025.00146},
publisher = {IEEE Computer Society},
address = {Los Alamitos, CA, USA},
month =May}

@article{nps,
author = {Reichheld, Frederick},
year = {2004},
month = {06},
pages = {46-54, 124},
title = {The One Number you Need to Grow},
volume = {81},
journal = {Harvard business review}
}

@misc{dynamicviz,
  year = "2019",
  title = "A system for dynamic slicing and program visualization",
  url = "https://dspace.mit.edu/handle/1721.1/121764",
  lastaccessed = "May 9, 2025",
}

@misc{whatisgdpr,
  title =        "What is GDPR?",
  url =          "https://gdpr.eu/what-is-gdpr/",
  lastaccessed = "May 28, 2024",
}

@misc{penalties,
  year = "2018",
  title = "GDPR penalties.",
  url = "https://gdpr-info.eu/issues/fines-penalties/",
  lastaccessed = "April 23, 2024",
}

@misc{wala,
  title = "WALA",
  url = "https://github.com/wala/WALA",
  lastaccessed = "June 18, 2024",
}

@misc{pbd,
  year =         "2009",
  author =       "Ann Cavoukian",
  title =        "Privacy by design: The 7 foundational principles",
  url =         "https://privacy.ucsc.edu/resources/privacy-by-design---foundational-principles.pdf",
  lastaccessed = "March 6, 2025",
}

@misc{dotgraph,
  title = "DotGraph",
  url = "https://www.sable.mcgill.ca/soot/doc/soot/util/dot/DotGraph.html",
  lastaccessed = "June 18, 2024",
}

@misc{cra,
    year = "2022",
    title = "Cyber Resilience Act",
    url = "https://digital-strategy.ec.europa.eu/en/library/cyber-resilience-act",
    lastaccessed = "April 23, 2024"
}

@misc{art13,
  year =         "2018",
  title =        "GDPR Article 13",
  url =          "https://gdpr-info.eu/art-13-gdpr/",
  lastaccessed = "April 23, 2024",
}

@misc{art25,
  year =         "2018",
  title =        "GDPR Article 25",
  url =          "https://gdpr-info.eu/art-25-gdpr/",
  lastaccessed = "June 16, 2024",
}

@misc{art32,
  year =         "2018",
  title =        "GDPR Article 32",
  url =          "https://gdpr-info.eu/art-32-gdpr/",
  lastaccessed = "May 26, 2025",
}

@misc{art4,
  year =         "2018",
  title =        "GDPR Article 4",
  url =          "https://gdpr-info.eu/art-4-gdpr/",
  lastaccessed = "Jan 29, 2025",
}

@misc{droidbench,
  year =         "2018",
  title =        "DroidBench",
  url =          "https://github.com/secure-software-engineering/DroidBench/tree/master",
  lastaccessed = "June 16, 2024",
}

@misc{art35,
  year =         "2018",
  title =        "GDPR Article 35",
  url =          "https://gdpr-info.eu/art-35-gdpr/",
  lastaccessed = "April 23, 2024",
}

@misc{googlechecks,
  year = "2022",
  author = "Nia Castelly and Fergus Hurley",
  title = "Introducing Checks:
simplifying privacy for app developers - Google:
The Keyword.",
  url = "https://blog.google/technology/area-120/checks/",
  lastaccessed = "April 23, 2024",
}

@misc{dpia,
  year =         "2018",
  title =        "Information Commissioner’s Office. 2018. Data Protection Impact Assessments (DPIAs).",
  url =          "https://ico.org.uk/for-organisations/guide-to-data-protection/guide-to-the-general-data-protection-regulation-gdpr/data-protection-impact-assessments-dpias/",
  lastaccessed = "May 13, 2024",
}

@INPROCEEDINGS{6979855,
  author={Moonsamy, Veelasha and Batten, Lynn},
  booktitle={2014 International Symposium on Information Theory and its Applications}, 
  title={Android applications: Data leaks via advertising libraries}, 
  year={2014},
  volume={},
  number={},
  pages={314-317},
  doi={}}

@ARTICLE{cognitiveload,
  author={Yoghourdjian, Vahan and Yang, Yalong and Dwyer, Tim and Lawrence, Lee and Wybrow, Michael and Marriott, Kim},
  journal={IEEE Transactions on Visualization and Computer Graphics}, 
  title={Scalability of Network Visualisation from a Cognitive Load Perspective}, 
  year={2021},
  volume={27},
  number={2},
  pages={1677-1687},
  doi={10.1109/TVCG.2020.3030459}}

@article{asej_price,
  author  = {Khedkar, Mugdha and Kumar Mondal, Ambuj and Bodden, Eric},
  title   = {A Study of Privacy-Related Data Collected by Android Apps},
  journal = {Automated Software Engineering},
  year    = {2026},
  volume  = {33},
  number  = {2},
  pages   = {45},
  doi     = {10.1007/s10515-025-00589-3},
  url     = {https://doi.org/10.1007/s10515-025-00589-3},
  issn    = {1573-7535}
}

@inproceedings{khedkarmobilesoft,
author = {Khedkar, Mugdha and Bodden, Eric},
title = {Toward an Android Static Analysis Approach for Data Protection},
year = {2024},
isbn = {9798400705946},
publisher = {Association for Computing Machinery},
address = {New York, NY, USA},
url = {https://doi.org/10.1145/3647632.3651389},
doi = {10.1145/3647632.3651389},
booktitle = {Proceedings of the IEEE/ACM 11th International Conference on Mobile Software Engineering and Systems},
pages = {65–68},
numpages = {4},
location = {Lisbon, Portugal},
series = {MOBILESoft '24}
}

@misc{datalabels,
      title={Unpacking Privacy Labels: A Measurement and Developer Perspective on Google's Data Safety Section}, 
      author={Rishabh Khandelwal and Asmit Nayak and Paul Chung and Kassem Fawaz},
      year={2023},
      eprint={2306.08111},
      archivePrefix={arXiv},
      primaryClass={cs.CY}
}

@INPROCEEDINGS{userstudy1,
  author={Berander, P.},
  booktitle={Proceedings. 2004 International Symposium on Empirical Software Engineering, 2004. ISESE '04.}, 
  title={Using students as subjects in requirements prioritization}, 
  year={2004},
  volume={},
  number={},
  pages={167-176},
  doi={10.1109/ISESE.2004.1334904}}

@INPROCEEDINGS{userstudy2,
  author={Salman, Iflaah and Misirli, Ayse Tosun and Juristo, Natalia},
  booktitle={2015 IEEE/ACM 37th IEEE International Conference on Software Engineering}, 
  title={Are Students Representatives of Professionals in Software Engineering Experiments?}, 
  year={2015},
  volume={1},
  number={},
  pages={666-676},
  doi={10.1109/ICSE.2015.82}}

@inproceedings{userstudy3,
author = {Svahnberg, Mikael and Aurum, Ayb\"{u}ke and Wohlin, Claes},
title = {Using students as subjects - an empirical evaluation},
year = {2008},
isbn = {9781595939715},
publisher = {Association for Computing Machinery},
address = {New York, NY, USA},
url = {https://doi.org/10.1145/1414004.1414055},
doi = {10.1145/1414004.1414055},
booktitle = {Proceedings of the Second ACM-IEEE International Symposium on Empirical Software Engineering and Measurement},
pages = {288–290},
numpages = {3},
location = {Kaiserslautern, Germany},
series = {ESEM '08}
}

@inproceedings{soot,
author = {Vall\'{e}e-Rai, Raja and Co, Phong and Gagnon, Etienne and Hendren, Laurie and Lam, Patrick and Sundaresan, Vijay},
title = {Soot - a Java Bytecode Optimization Framework},
year = {1999},
publisher = {IBM Press},
booktitle = {Proceedings of the 1999 Conference of the Centre for Advanced Studies on Collaborative Research},
pages = {13},
location = {Mississauga, Ontario, Canada},
series = {CASCON '99}
}

@ARTICLE{Weiser,
  author={Weiser, Mark},
  journal={IEEE Transactions on Software Engineering}, 
  title={Program Slicing}, 
  year={1984},
  volume={SE-10},
  number={4},
  pages={352-357},
  doi={10.1109/TSE.1984.5010248}}

@article{Weiser1,
author = {Weiser, Mark},
title = {Programmers use slices when debugging},
year = {1982},
issue_date = {July 1982},
publisher = {Association for Computing Machinery},
address = {New York, NY, USA},
volume = {25},
number = {7},
issn = {0001-0782},
url = {https://doi.org/10.1145/358557.358577},
doi = {10.1145/358557.358577},
journal = {Commun. ACM},
month = {jul},
pages = {446–452},
numpages = {7}
}

@inproceedings{ppviolationappcode,
author = {Slavin, Rocky and Wang, Xiaoyin and Hosseini, Mitra Bokaei and Hester, James and Krishnan, Ram and Bhatia, Jaspreet and Breaux, Travis D. and Niu, Jianwei},
title = {Toward a framework for detecting privacy policy violations in android application code},
year = {2016},
isbn = {9781450339001},
publisher = {Association for Computing Machinery},
address = {New York, NY, USA},
url = {https://doi.org/10.1145/2884781.2884855},
doi = {10.1145/2884781.2884855},
booktitle = {Proceedings of the 38th International Conference on Software Engineering},
pages = {25–36},
numpages = {12},
location = {Austin, Texas},
series = {ICSE '16}
}

@inproceedings{franke2024,
author = {Franke, Lucas and Liang, Huayu and Farzanehpour, Sahar and Brantly, Aaron and Davis, James C. and Brown, Chris},
title = {An Exploratory Mixed-methods Study on General Data Protection Regulation (GDPR) Compliance in Open-Source Software},
year = {2024},
isbn = {9798400710476},
publisher = {Association for Computing Machinery},
address = {New York, NY, USA},
url = {https://doi.org/10.1145/3674805.3686692},
doi = {10.1145/3674805.3686692},
booktitle = {Proceedings of the 18th ACM/IEEE International Symposium on Empirical Software Engineering and Measurement},
pages = {325–336},
numpages = {12},
location = {Barcelona, Spain},
series = {ESEM '24}
}

@inproceedings{mandoline,
  author       = {Khaled Ahmed and
                  Mieszko Lis and
                  Julia Rubin},
  title        = {Mandoline: Dynamic Slicing of Android Applications with Trace-Based
                  Alias Analysis},
  booktitle    = {14th {IEEE} Conference on Software Testing, Verification and Validation,
                  {ICST} 2021, Porto de Galinhas, Brazil, April 12-16, 2021},
  pages        = {105--115},
  publisher    = {{IEEE}},
  year         = {2021},
  url          = {https://doi.org/10.1109/ICST49551.2021.00022},
  doi          = {10.1109/ICST49551.2021.00022},
  bibsource    = {dblp computer science bibliography, https://dblp.org}
}

@inproceedings{dynamicslicing,
author = {Azim, Tanzirul and Alavi, Arash and Neamtiu, Iulian and Gupta, Rajiv},
title = {Dynamic slicing for Android},
year = {2019},
publisher = {IEEE Press},
url = {https://doi.org/10.1109/ICSE.2019.00118},
doi = {10.1109/ICSE.2019.00118},
booktitle = {Proceedings of the 41st International Conference on Software Engineering},
pages = {1154–1164},
numpages = {11},
location = {Montreal, Quebec, Canada},
series = {ICSE '19}
}

@article{surveyslicing,
author = {Xu, Baowen and Qian, Ju and Zhang, Xiaofang and Wu, Zhongqiang and Chen, Lin},
title = {A brief survey of program slicing},
year = {2005},
issue_date = {March 2005},
publisher = {Association for Computing Machinery},
address = {New York, NY, USA},
volume = {30},
number = {2},
issn = {0163-5948},
url = {https://doi.org/10.1145/1050849.1050865},
doi = {10.1145/1050849.1050865},
journal = {SIGSOFT Softw. Eng. Notes},
month = {mar},
pages = {1–36},
numpages = {36}
}

@article{matcha,
   title={Matcha: An IDE Plugin for Creating Accurate Privacy Nutrition Labels},
   volume={8},
   ISSN={2474-9567},
   url={http://dx.doi.org/10.1145/3643544},
   DOI={10.1145/3643544},
   number={1},
   journal={Proceedings of the ACM on Interactive, Mobile, Wearable and Ubiquitous Technologies},
   publisher={Association for Computing Machinery (ACM)},
   author={Li, Tianshi and Cranor, Lorrie Faith and Agarwal, Yuvraj and Hong, Jason I.},
   year={2024},
   month=mar, pages={1–38} }

@ARTICLE {analyticslibrarysource,
author = {D. Caputo and F. Pagano and G. Bottino and L. Verderame and A. Merlo},
journal = {IEEE Transactions on Dependable and Secure Computing},
title = {You Can\'t Always Get What You Want: Towards User-Controlled Privacy on Android},
year = {2023},
volume = {20},
number = {02},
issn = {1941-0018},
pages = {975-987},
doi = {10.1109/TDSC.2022.3146020},
publisher = {IEEE Computer Society},
address = {Los Alamitos, CA, USA},
month = {mar}
}

@inproceedings{tpl,
author = {Backes, Michael and Bugiel, Sven and Derr, Erik},
title = {Reliable Third-Party Library Detection in Android and its Security Applications},
year = {2016},
isbn = {9781450341394},
publisher = {Association for Computing Machinery},
address = {New York, NY, USA},
url = {https://doi.org/10.1145/2976749.2978333},
doi = {10.1145/2976749.2978333},
booktitle = {Proceedings of the 2016 ACM SIGSAC Conference on Computer and Communications Security},
pages = {356–367},
numpages = {12},
location = {Vienna, Austria},
series = {CCS '16}
}

@article{taintbench, title={TaintBench: Automatic real-world malware benchmarking of Android taint analyses}, DOI={10.1007/s10664-021-10013-5}, journal={Empirical Software Engineering}, author={Luo, Linghui and Pauck, Felix and Piskachev, Goran and Benz, Manuel and Pashchenko, Ivan and Mory, Martin and Bodden, Eric and Hermann, Ben and Massacci, Fabio}, year={2021} }

@article{hashingpitfalls,
  author={Demir, Levent and Kumar, Amrit and Cunche, Mathieu and Lauradoux, Cédric},
  journal={IEEE Communications Surveys \& Tutorials}, 
  title={The Pitfalls of Hashing for Privacy}, 
  year={2018},
  volume={20},
  number={1},
  pages={551-565},
  doi={10.1109/COMST.2017.2747598}
}

@techreport{enisa2,
	author = {{Konstantinos Limniotis, Marit Hansen}},
	title = {Recommendations on shaping technology according to GDPR provisions - An overview on data pseudonymisation},
	year = {2019},
	institution = {European Union Agency for Cybersecurity (ENISA)},
    url = {https://www.enisa.europa.eu/publications/recommendations-on-shaping-technology-according-to-gdpr-provisions},
}

@INPROCEEDINGS{privacypolicytrust,
  author={Yu, Le and Luo, Xiapu and Liu, Xule and Zhang, Tao},
  booktitle={2016 46th Annual IEEE/IFIP International Conference on Dependable Systems and Networks (DSN)}, 
  title={Can We Trust the Privacy Policies of Android Apps?}, 
  year={2016},
  volume={},
  number={},
  pages={538-549},
  doi={10.1109/DSN.2016.55}}

@inbook{automatedriskanalysis,
title = "Automated Analysis of Privacy Requirements for Mobile Apps",
author = "Sebastian Zimmeck and Ziqi Wang and Lieyong Zou and Roger Iyengar and Bin Liu and Florian Schaub and Shomir Wilson and Norman Sadeh and Bellovin, {Steven M.} and Joel Reidenberg",
year = "2017",
doi = "10.14722/ndss.2017.23034",
language = "English (US)",
isbn = "1-891562-46-0",
series = "Proceedings 2017 Network and Distributed System Security Symposium",
publisher = "Korea Society of Internet Information",
booktitle = "Proceedings 2017 Network and Distributed System Security Symposium",
address = "Korea, Republic of",

}

@inproceedings{guileak,
author = {Wang, Xiaoyin and Qin, Xue and Hosseini, Mitra Bokaei and Slavin, Rocky and Breaux, Travis D. and Niu, Jianwei},
title = {GUILeak: Tracing Privacy Policy Claims on User Input Data for Android Applications},
year = {2018},
isbn = {9781450356381},
publisher = {Association for Computing Machinery},
address = {New York, NY, USA},
url = {https://doi.org/10.1145/3180155.3180196},
doi = {10.1145/3180155.3180196},
booktitle = {Proceedings of the 40th International Conference on Software Engineering},
pages = {37–47},
numpages = {11},
location = {Gothenburg, Sweden},
series = {ICSE '18}
}

@inproceedings{slicingSDG,
author = {Horwitz, S. and Reps, T. and Binkley, D.},
title = {Interprocedural Slicing Using Dependence Graphs},
year = {1988},
isbn = {0897912691},
publisher = {Association for Computing Machinery},
address = {New York, NY, USA},
url = {https://doi.org/10.1145/53990.53994},
doi = {10.1145/53990.53994},
booktitle = {Proceedings of the ACM SIGPLAN 1988 Conference on Programming Language Design and Implementation},
pages = {35–46},
numpages = {12},
location = {Atlanta, Georgia, USA},
series = {PLDI '88}
}

@INPROCEEDINGS{jicer,
  author={Pauck, Felix and Wehrheim, Heike},
  booktitle={2021 IEEE 21st International Working Conference on Source Code Analysis and Manipulation (SCAM)}, 
  title={Jicer: Simplifying Cooperative Android App Analysis Tasks}, 
  year={2021},
  volume={},
  number={},
  pages={187-197},
  doi={10.1109/SCAM52516.2021.00031}}

@article{thinslicing,
author = {Sridharan, Manu and Fink, Stephen J. and Bodik, Rastislav},
title = {Thin slicing},
year = {2007},
issue_date = {June 2007},
publisher = {Association for Computing Machinery},
address = {New York, NY, USA},
volume = {42},
number = {6},
issn = {0362-1340},
url = {https://doi.org/10.1145/1273442.1250748},
doi = {10.1145/1273442.1250748},
journal = {SIGPLAN Not.},
month = {jun},
pages = {112–122},
numpages = {11}
}

@misc{atpchecker,
  doi = {10.48550/ARXIV.2301.12348},
  url = {https://arxiv.org/abs/2301.12348},
  author = {Zhao, Kaifa and Zhan, Xian and Yu, Le and Zhou, Shiyao and Zhou, Hao and Luo, Xiapu and Wang, Haoyu and Liu, Yepang},
  title = {Demystifying Privacy Policy of Third-Party Libraries in Mobile Apps},
  publisher = {arXiv},
  year = {2023},
  copyright = {arXiv.org perpetual, non-exclusive license}
}

@inproceedings{privflow,
author = {Tang, Feiyang and \O{}stvold, Bjarte M.},
title = {Assessing Software Privacy Using the Privacy Flow-Graph},
year = {2022},
isbn = {9781450394574},
publisher = {Association for Computing Machinery},
address = {New York, NY, USA},
url = {https://doi.org/10.1145/3549035.3561185},
doi = {10.1145/3549035.3561185},
booktitle = {Proceedings of the 1st International Workshop on Mining Software Repositories Applications for Privacy and Security},
pages = {7–15},
numpages = {9},
location = {Singapore, Singapore},
series = {MSR4P\&S 2022}
}

@inproceedings{mudflow,
author = {Avdiienko, Vitalii and Kuznetsov, Konstantin and Gorla, Alessandra and Zeller, Andreas and Arzt, Steven and Rasthofer, Siegfried and Bodden, Eric},
title = {Mining Apps for Abnormal Usage of Sensitive Data},
year = {2015},
isbn = {9781479919345},
publisher = {IEEE Press},
booktitle = {Proceedings of the 37th International Conference on Software Engineering - Volume 1},
pages = {426–436},
numpages = {11},
location = {Florence, Italy},
series = {ICSE '15}
}

@INPROCEEDINGS{PTPDroid,
  author={Tan, Zeya and Song, Wei},
  booktitle={2023 IEEE/ACM 45th International Conference on Software Engineering (ICSE)}, 
  title={PTPDroid: Detecting Violated User Privacy Disclosures to Third-Parties of Android Apps}, 
  year={2023},
  volume={},
  number={},
  pages={473-485},
  doi={10.1109/ICSE48619.2023.00050}}

@inproceedings{flowdroid,
author = {Arzt, Steven and Rasthofer, Siegfried and Fritz, Christian and Bodden, Eric and Bartel, Alexandre and Klein, Jacques and Le Traon, Yves and Octeau, Damien and McDaniel, Patrick},
title = {FlowDroid: Precise Context, Flow, Field, Object-Sensitive and Lifecycle-Aware Taint Analysis for Android Apps},
year = {2014},
isbn = {9781450327848},
publisher = {Association for Computing Machinery},
address = {New York, NY, USA},
url = {https://doi.org/10.1145/2594291.2594299},
doi = {10.1145/2594291.2594299},
booktitle = {Proceedings of the 35th ACM SIGPLAN Conference on Programming Language Design and Implementation},
pages = {259–269},
numpages = {11},
location = {Edinburgh, United Kingdom},
series = {PLDI '14}
}
